\documentclass[aps,twocolumn,pra,superscript,floatfix,superscriptaddress,showpacs,footinbib,reprint]{revtex4-2}
\usepackage{amssymb}

\usepackage[pdftex]{graphicx}
\usepackage{dcolumn}% Align table columns on decimal point
\usepackage{bm}% bold math
\usepackage{amsmath}
\usepackage{nicefrac}
\usepackage{array}
\usepackage{color}
\usepackage{float}
\usepackage{subfigure}
\usepackage{dsfont}
\usepackage{txfonts}
\usepackage{wasysym}
\usepackage{multirow}
\usepackage{sidecap}
\usepackage{xcolor,cancel}
\usepackage[plainpages=false,pdfpagelabels,colorlinks=true,linkcolor=red,urlcolor=blue,
citecolor=blue,pdftitle={BE15025-resub},pdfauthor={},pdfdisplaydoctitle=true]{hyperref}
\usepackage[normalem]{ulem}

\newcommand{\+}{\dagger}

\newcommand{\dn}{\downarrow}

\newcommand{\e}{\varepsilon}

\newcommand{\hyb}{\mathrm{hyb}}

\newcommand{\up}{\uparrow}

\newcommand{\imp}{{\rm imp}}

\renewcommand{\Im}{{\rm Im \,}}

\newcommand{\HC}{{ \rm H.c.}}

\begin{document}

\title{Quantum impurity with \nicefrac{2}{3} local moment in 1D quantum wires: an NRG study}

\author{P.~A.~Almeida}
\affiliation{Instituto de F\'isica, Universidade Federal de Uberl\^andia, 
Uberl\^andia, Minas Gerais, 38400-902, Brazil}

\author{M.~A.~Manya}
\affiliation{Instituto de F\'isica, Universidade Federal Fluminense,
Av. Litor\^anea s/N, 24210-340, Niter\'oi, RJ, Brazil}

\author{M.~S.~Figueira}
\affiliation{Instituto de F\'isica, Universidade Federal Fluminense,
Av. Litor\^anea s/N, 24210-340, Niter\'oi, RJ, Brazil}

\author{S.~E.~Ulloa}
\affiliation{Department of Physics and Astronomy and Nanoscale and Quantum Phenomena Institute,
Ohio University, Athens, Ohio 45701-2979, USA}

\author{E.~V.~Anda}
\affiliation{Departamento de F\'isica, Pontif\'icia Universidade Cat\'olica 
do Rio de Janeiro (PUC-Rio), Rio de Janeiro, Rio de Janeiro, 22453-900, Brazil}

\author{G. B. Martins}
\affiliation{Instituto de F\'isica, Universidade Federal de Uberl\^andia, 
Uberl\^andia, Minas Gerais, 38400-902, Brazil}
\email[Corresponding author: ]{gbmartins@ufu.br}

\date{\today}
\begin{abstract}
We study a Kondo state that is strongly influenced
by its proximity to an $\omega^{-1/2}$ singularity in the metallic 
host density of states. This singularity occurs at the bottom of the band of a one-dimensional  
chain, for example. We first analyze the non-interacting system: A
resonant state $\e_d$, located close to the band singularity, suffers 
a strong `renormalization', such that a bound state (Dirac delta function) 
is created below the bottom of the band in addition to a resonance in the continuum. 
When $\e_d$ is positioned right at the singularity, the spectral weight of the bound state is
$\nicefrac{2}{3}$, irrespective of its coupling to the 
conduction electrons. The interacting system is modeled using the Single Impurity
Anderson Model, which is then solved using the Numerical
Renormalization Group method. We observe that the Hubbard interaction causes 
the bound state to suffer a series of transformations, including level splitting, 
transfer of spectral weight, appearance of a spectral discontinuity, changes in 
binding energy (the lowest state moves farther away from the bottom of the band), 
and development of a finite width. When $\e_d$ is away from the 
singularity and in the intermediate valence regime, the impurity occupancy is lower. 
As $\e_d$ moves closer to the singularity, the system partially recovers Kondo regime
properties, i.e., higher occupancy and lower Kondo temperature $T_K$. The impurity 
thermodynamic properties show that the local moment fixed point is also 
strongly affected by the existence of the bound state. 
When $\e_d$ is close to the singularity, the  local moment 
fixed point becomes impervious to charge fluctuations (caused by  
bringing $\e_d$ close to the Fermi energy), in contrast to the local moment 
suppression that occurs when $\e_d$ is away from the singularity.
We also discuss an experimental implementation that shows similar results to the quantum wire,  
if the impurity's metallic host is an armchair graphene nanoribbon. 

\end{abstract}

\maketitle

\section{Introduction}
The Kondo effect~\cite{Hewson1993} has been extensively studied, 
both theoretically~\cite{Krishna-murthy1980} and 
experimentally~\cite{Madhavan1998,Goldhaber-Gordon1998}, and it is considered 
one of the pillars of many-body physics~\cite{Coleman2015}. 
It is simulated by a quantum impurity coupled to a non-interacting Fermi 
sea, through a model that may include charge fluctuations, 
resulting in the well-known Single Impurity Anderson model (SIAM)~\cite{Anderson1961}, 
or through a model that accounts only for the strong-coupling fixed point, where 
just spin fluctuations are relevant~\cite{Schrieffer1966} -- the so-called 
Kondo model~\cite{Kondo1964}. The Numerical Renormalization Group (NRG) method was developed 
in the 1970s~\cite{Wilson1975,Krishna-murthy1980,Wilson1983,Bulla2008},  
and it is uniquely able to tackle the Kondo problem.  
To this day, it is among the most popular techniques to deal with this 
fascinating problem.
The main properties of the Kondo state, the quenching of the 
impurity magnetic moment, universal temperature scaling, and  
existence of renormalization fixed points, are readily obtained when 
considering a featureless (`flat') density of states (DOS) of the host around the 
Fermi energy $E_F$, where most of the important physics occurs. This may be 
called a `traditional Kondo effect'. 
Things become more interesting, possibly including non-Fermi liquid physics~\cite{Vojta2002}, 
when the host DOS behaves like $\rho(\omega) = \vert\omega\vert^r$ 
\emph{at}, or near, the Fermi energy. For $r>0$, the DOS vanishes at $\omega=0$ and 
the band is said to have a pseudogap. Many theoretical works have analyzed 
the Kondo model (no charge fluctuations) for bands presenting a 
pseudogap~\cite{Withoff1990,Borkowski1992,KanChen1995,Ingersent1996,Bulla1997,
Gonzalez-Buxton1998,Bulla2000,Ingersent2002,Vojta2002,Dias2009}, while much less work has been devoted 
to the $r<0$ case, i.e., when there is a divergent DOS (singularity) \emph{at} the Fermi 
energy~\cite{Vojta2002,Zitko2009a,Mitchell2013a,Mitchell2013b}. 
Even fewer works have discussed~\cite{Zhuravlev2009,Zhuravlev2011,Zitko2011,Shchadilova2014,Zitko2016,Wong2016} 
how a singularity \emph{close} to the Fermi energy, generating 
high particle-hole asymmetry, modifies 
the Kondo state. Recent work~\cite{Agarwala2016} discusses a Kondo state 
where the impurity orbital level is resonant with a singularity 
at the bottom of the band (a situation that occurs for 
a one-dimensional (1D) lattice, nanotubes~\cite{Charlier2007}, and 
nanoribbons~\cite{Wakabayashi2010}), while the Fermi energy is slightly  
above the singularity, with very interesting results.  

\begin{figure}[h]
\includegraphics[width=1.0\columnwidth]{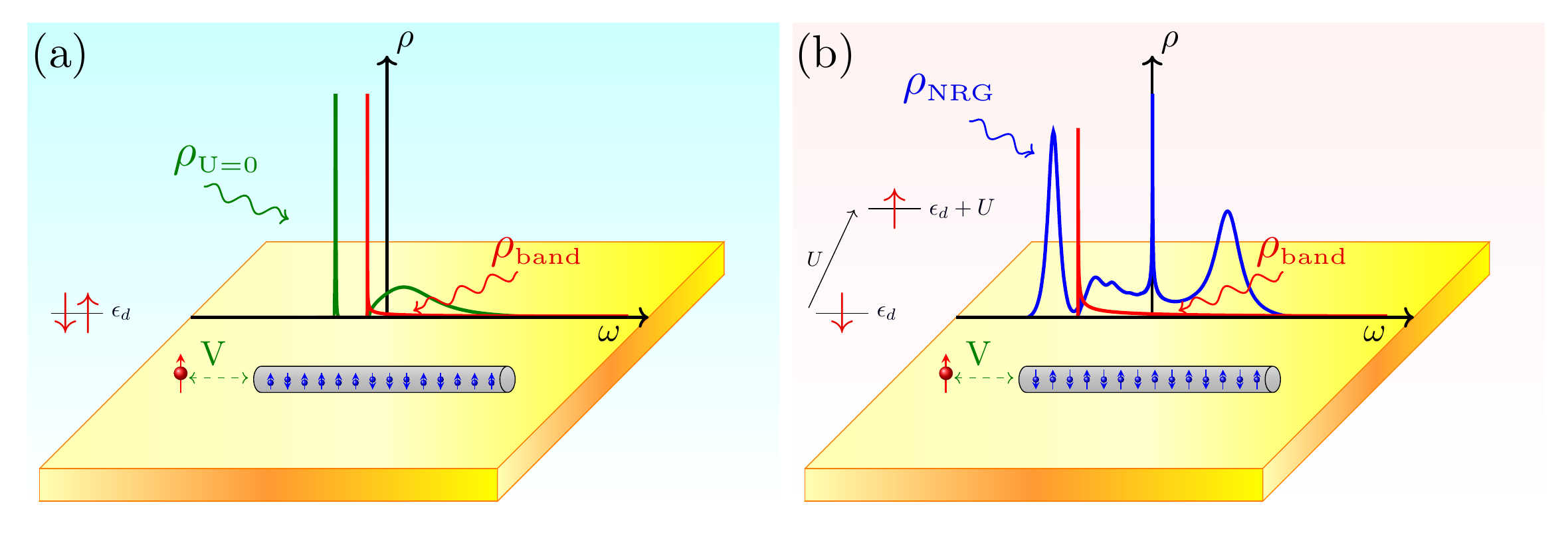}
	\caption{\label{fig1} Comparison of non-interacting and interacting results. 
	(a) The red curve shows host DOS with 
	$\omega^{\nicefrac{-1}{2}}$ singularity at the band bottom, while the green curve 
	presents the DOS for a non-interacting 
	impurity, showing a well-defined bound state below the continuum when 
	$\e_d$ is resonant with the band singularity. (b) 
	Same as in (a), but now for an interacting impurity ($U \ne 0$). 
	The bound state acquires a finite width, while a well-developed 
	Kondo peak appears at the Fermi energy (see Appendix D for a critical 
	discussion concerning the width of the bound state).
	} 
\end{figure}

In this work, we revisit a situation similar to the last system~\cite{Agarwala2016}, 
using the NRG method. To better understand the system here, we first study the \emph{non-interacting} 
regime. In the Appendix, we discuss in detail what happens when a non-interacting 
resonant level (RL), generically called an impurity, is placed near (or \emph{at}) 
the $\omega^{\nicefrac{-1}{2}}$
singularity that occurs naturally in a 1D quantum wire host (see Fig.~\ref{fig1}(a) 
for an illustration). We show that the non-interacting RL spectral 
function, $\rho_{U=0}$ [green curve in Fig.~\ref{fig1}(a)], exhibits a 
bound state below the bottom of the band DOS, $\rho_{band}$ (red curve). 
The bound state has the properties 
of a Dirac delta function (see Appendix) and carries a spectral weight 
that depends on the coupling of the RL to the band and 
its distance to the singularity in $\rho_{band}$ 
[visible as a sharp red peak in Fig.~\ref{fig1}(a)]. 
Our results show a couple of interesting features of this deceptively simple non-interacting 
problem. First, there is a bound state even when the 
RL is positioned close to the center of the band. In that case, 
the bound state appears exactly at the edge of the band 
but carries negligible spectral weight, even for moderate coupling to the 
band (see discussion around Fig.~\ref{fig13} in the Appendix). Second, if the 
RL is positioned at the singularity, 
the bound state carries a spectral weight that is exactly $\nicefrac{2}{3}$~\cite{SOI}, 
irrespective of the coupling-strength of the RL to the band. 

Once the Hubbard interaction $U$ is turned on [see Fig.~\ref{fig1}(b)], we find that the 
usual Kondo profile of the impurity spectral function [$\rho_{NRG}$, blue curve in 
Fig.~\ref{fig1}(b)] is modified. Indeed, the singularity strongly distorts the impurity's 
lower Coulomb Blockade peak (CBP), which is now composed of a \emph{broadened} bound state and a 
series of peaks. 
In addition, the Kondo temperature and the impurity occupancy 
are strongly affected when $\e_d$ is close to the band singularity. 
Both quantities tend to values closer to those 
fully in the Kondo regime (higher occupancy and lower $T_K$) 
even when the system is in an intermediate valence regime. This 
`reentrant' Kondo regime can be also observed at temperatures around the local 
moment (LM) fixed point. Indeed, the magnetic susceptibility 
in the intermediate valence regime takes values similar 
to those in the Kondo regime at temperatures associated to the LM fixed point. 
This behavior, also visible in the NRG energy flow, is clearly associated 
to the existence of the bound state. We present NRG results and analysis to 
explore these interesting regimes in detail below.

The paper is organized as follows: In Sec.~\ref{sec-mod}, we present the Hamiltonian 
for the system to be analyzed, while Sec.~\ref{sec-NRG} presents the NRG 
results. This section is divided into three parts: 
Sec.~\ref{sec-spectral} presents the dependence of the impurity spectral 
function (and charge occupancy) on the proximity of the impurity 
orbital level to the singularity at the bottom of the band. 
Subsection \ref{sec-correl} tracks how the bound state present in the 
non-interacting problem ($U=0$) is affected by the introduction of 
correlations (finite $U$). In Sec.~\ref{sec-thermo}, we 
analyze the impurity susceptibility, as well as the impurity entropy, 
and verify that the correlated states caused by the presence of the singularity 
have a strong influence on the impurity properties 
close to the LM fixed point. 
The thermodynamic results are interpreted through an analysis 
of the NRG energy flow. Sec.~\ref{sec-nanoribbon} presents 
results for an experimentally accessible system, where these effects 
could be observed, viz., an $N=3$ armchair 
graphene nanoribbon. This system has two singularities, at the bottom of the 
conduction and valence bands, that have an $\omega^{\nicefrac{-1}{2}}$-dependence, equal 
to that in a quantum wire system. 
Sec.~\ref{sec-conc} presents a discussion of the results and our conclusions. 
Finally, as mentioned above, in Appendices~\ref{appA} to~\ref{appC}, we analyze in detail 
the non-interacting system when the RL is close to the singularity, while in 
Appendix~\ref{appD} we study the interacting (NRG) impurity spectral 
function to ascertain that the results in Sec.~\ref{sec-spectral} do not contain 
numerical artifacts.

\section{Model and Hamiltonian}\label{sec-mod}
In the following, we analyze the Kondo effect of an impurity 
coupled to a 1D quantum wire. 
We start with the quantum wire Hamiltonian 

\begin{eqnarray}\label{h1D}
%	H_{\rm wire}&=\sum_{k,\sigma}  \left(-2t \cos k  - \mu \right)c^\dagger_{k\sigma}c_{k\sigma},
	H_{\text {wire }}=\displaystyle{\sum_{k, \sigma}}(-2 t \cos k-\mu) c_{k \sigma}^{\dagger} c_{k \sigma}
\end{eqnarray}
where $c^\dagger_{k\sigma}$ creates an electron with wave 
vector $k$ and spin $\sigma=\up,\dn$, while $t=0.5$ is the 
nearest-neighbor hopping in the tight-binding chain (thus, 
$D=1$, the half bandwidth, is our unit of energy), and  
$\mu$ is the chemical potential. The Fermi energy, for different values 
of $\mu$, is always set at zero ($E_F=0$). 

To study the Kondo state in this system, the wire is coupled to an Anderson impurity, 
which is modeled as  
\begin{eqnarray}\label{eq:himp}
	H_\imp=\sum_{\sigma} \e_d n_\sigma + Un_\up n_\dn, 
\end{eqnarray}
where $d^\+_{\sigma}$ ($d_{\sigma}$) creates (annihilates) an electron with orbital 
energy $\e_d$ and spin $\sigma=\uparrow, \downarrow$, $n_{\sigma}=d^\+_{\sigma}d_{\sigma}$, and 
$U$ represents the Coulomb repulsion. 
The hybridization between the impurity and the conduction electrons is given by
\begin{eqnarray}
	H_\hyb=\sum_{k\sigma}\left( V_k  d^{\dagger}_{\sigma} c_{k\sigma}+\HC \right),
\end{eqnarray}
where we consider the case of $V_k \equiv V$. The parameter 
$\Gamma=\pi V^2 \rho_{band}(E_F)$ determines the strength of the coupling of the impurity to the 
bath, where $\rho_{band}(E_F)$ is the host's DOS at the Fermi energy. To solve this problem, 
we use the well-known NRG Ljubljana open source code~\cite{zitko_rok}. 
For most of the calculations, we have used the discretization 
parameter $\Lambda=2.0$ and kept at least 5000 states at each iteration. We also employ the 
so-called z-trick~\cite{Campo2005} (with $z=0.0625$, $0.125$, $\ldots$, and $1.0$, i.e., $N_z=16$) to remove 
oscillations (artifacts) in the physical quantities. The Kondo temperature was obtained through 
Wilson's criterion~\cite{Hewson1993}, namely, the temperature for which the 
impurity susceptibility, multiplied by the temperature, reaches $0.07$. 
The thermodynamic quantities were calculated using the traditional single-shell approximation, 
while the dynamical quantities (spectral function) were calculated using the Density Matrix 
NRG approximation~\cite{Hofstetter2000}. Finally, the single particle calculations 
in the main text (and in Appendix~\ref{appD}) have used an imaginary part $\eta=10^{-6}$ to regularize the 
Green's function. 

\section{NRG results}\label{sec-NRG}

\subsection{Singularity effect on the impurity spectral function and charge occupancy}\label{sec-spectral}

As described in the literature~\cite{Zhuravlev2009,Zhuravlev2011,Zitko2011,
Shchadilova2014,Zitko2016,Wong2016}, the Kondo state  
for Anderson-type systems~\cite{Anderson1961} and highly asymmetric DOS 
(such as when the Fermi energy $E_F$ is close to a Van Hove singularity) 
strongly depends on model parameters. Indeed, our detailed analysis of the Kondo state 
for $E_F$ close to the 1D band-singularity~\cite{chempot} indicates that the 
impurity spectral function, impurity charge occupancy, and 
thermodynamic properties, are very sensitive to the interplay 
between $\e_d$, $U$, $V$, and $E_F$. In other words, small changes in the 
parameters, like the position of $E_F$ in relation to the singularity, 
strongly affect the Kondo state. 

\begin{figure}[h]
\includegraphics[width=1.0\columnwidth]{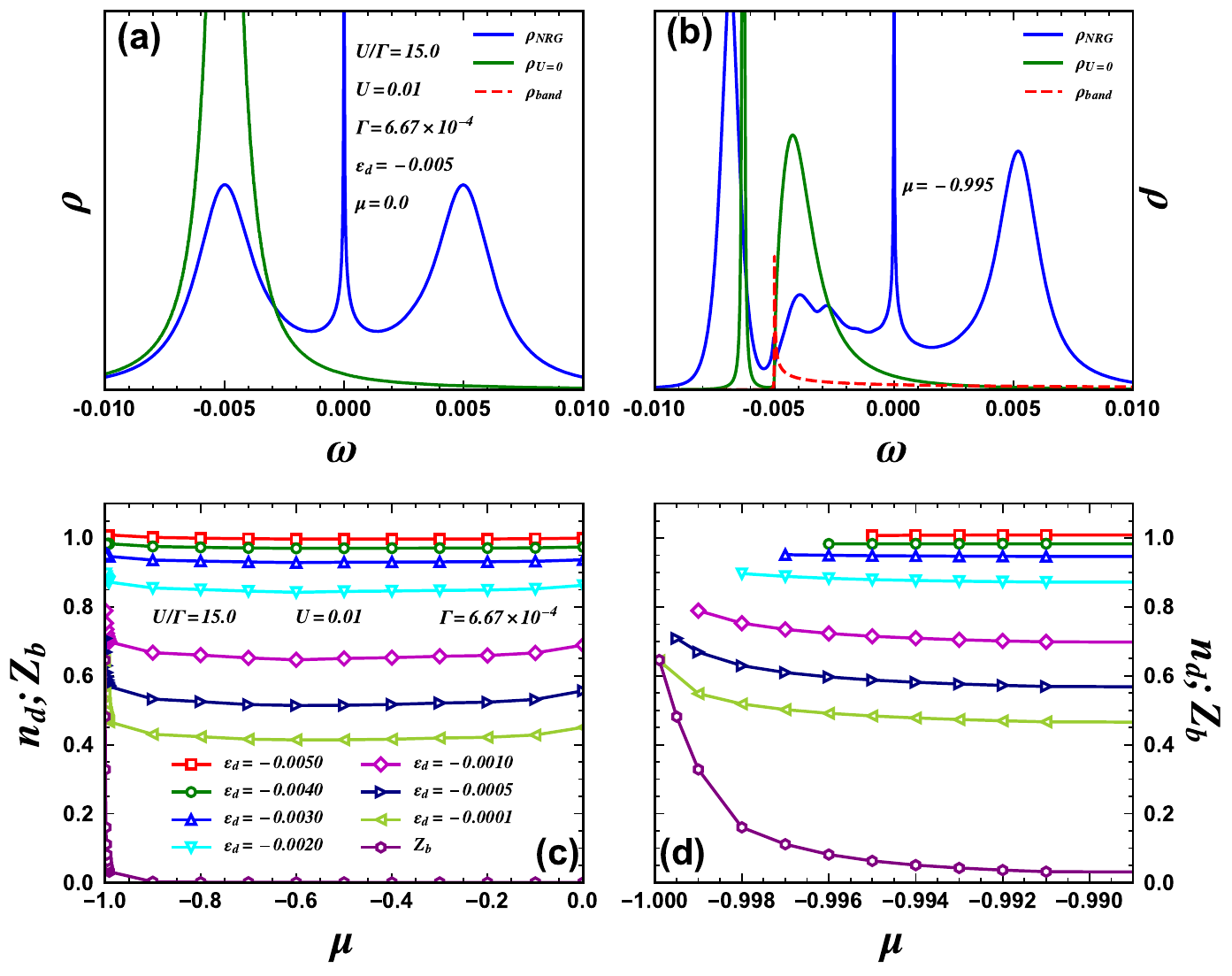}
	\caption{\label{fig2} DOS comparison and spectral weights for $U=0$ and $U=0.01$.
	(a) Impurity DOS for $U=0$ ($\rho_{U=0}$, green curve) 
	and for $\nicefrac{U}{\Gamma}=15$ ($\rho_{NRG}$, blue curve) 
	for $\mu=0$ (when $E_F$ is at the center of the 
	band, $\omega=0.0$), $U=0.01$ and $\e_d=\nicefrac{-U}{2}=-0.005$ 
	(notice dashed-red-curve close to 
	the horizontal axis, showing the band DOS, $\rho_{band}$). 
	(b) Same as in (a), but now with $E_F$ very close to the singularity 
	at the bottom of the band (depicted by the dashed red curve, $\rho_{band}$), 
	$\mu=-0.995$, $\e_d=-0.005$ (right at the singularity). 
	In (a) and (b), the $\Gamma$ value is the same for finite and vanishing $U$. 
	(c) Finite $U=0.01$ results for the evolution of 
	$n_d= \langle n_{d\uparrow} + n_{d\downarrow}\rangle$ as $E_F$ moves from the 
	center of the band ($\mu=0$, right side) to very near the singularity ($\mu=-1.0-\e_d$, 
	left side), for different 
	values of $\e_d$ (the leftmost value of $\mu$ places the RL exactly 
	at the singularity). The lowest (purple hexagons) curve with hexagons shows the bound state 
	spectral weight $Z_b$ (see Appendix), for $U=0$ and $\e_d=-0.0001$. 
	(d) Zoom of the left side of panel (c) highlighting the abrupt increase 
	in $n_d$. The lowest value of the chemical potential is 
	$\mu=-1.0 - \e_d$, thus the curves above do not cover the same $\mu$ interval. 
	} 
\end{figure}

To reveal the most interesting aspects of the Kondo state when $\e_d$ is at 
the singularity (and $E_F$ is close to the bottom of the band), 
we will contrast it to the Kondo state obtained when $E_F$ is exactly in the middle 
of the band ($\mu=0$), keeping all the other parameters equal. We take 
$\e_d=\nicefrac{-U}{2}$, thus, for $\mu=0$ the system is in the 
particle-hole-symmetric (PHS) point. These results are shown 
in Fig.~\ref{fig2}. Panels (a) and (b) show the impurity spectral function 
(green curve, for the non-interacting case, $\rho_{U=0}$, and  blue curve, 
for the interacting case, $\rho_{NRG}$) for 
$\mu=0$ and $\mu=-0.995$, respectively. The (red) dashed curves are the band DOS, 
$\rho_{band}$. Note that all DOS results are normalized so that their integrals 
over $\omega$ are $1$. The parameters, kept fixed for both calculations, 
are $\e_d=-0.005$, $U=0.01$, and $\Gamma=6.6667 \times 10^{-4}$ (thus 
$\nicefrac{U}{\Gamma}=15$). We have used $V$ values for both calculations 
($V=0.026$ and $V=0.0082$, for panels (a) and (b), respectively) such that 
$\Gamma$ does not vary. The only change from one calculation to the 
other is the PH asymmetry around $E_F$: no asymmetry in panel (a) 
and a very strong asymmetry in panel (b). Comparison of the 
$\rho_{NRG}$ results (blue curves) in panels (a) and (b) 
shows how strongly the singularity affects the impurity spectral density~\cite{edge}. 
Indeed, from a traditional PHS Kondo peak at $\mu=0$ [panel (a), blue curve], 
we move to a very rich impurity DOS 
when $\e_d$ is at the singularity, showing a series of peaks around the 
Fermi energy ($E_F=0.0$). The rightmost peak in $\rho_{NRG}$ (the upper CBP), 
farthest from the singularity, is the least affected, 
while the Kondo peak (around $\omega=0.0$) acquires a slight asymmetry. 
Notice that the RL results (green curve, $\rho_{U=0}$)
show that the singularity splits the 
non-interacting DOS into a Dirac delta-like bound state (below the bottom 
of the band) and a broad peak starting at the bottom of the 
band. As it turns out, this last peak becomes a superposition of three peaks 
in the continuum, while the bound state splits into two features below the band. 
One is very sharp, located at the band edge, and has very small spectral weight. 
The other, containing most of the spectral weight, is shifted to 
lower energy than the original bound state, and acquires a sizable finite width~\cite{finitewidth}. 
Thus, the interplay between the singularity and correlations results in very complex 
spectral behavior. This will be further analyzed in the next subsection, \ref{sec-correl}.  

Fig.~\ref{fig2}(c) shows how the impurity occupancy 
$n_d = \langle n_{d\uparrow} + n_{d\downarrow}\rangle$ 
(for $U=0.01$ and $\nicefrac{U}{\Gamma}=15$) varies when the Fermi energy moves from the 
center of the band ($\mu=0$) to close to the bottom of the band ($\mu = -1.0 - \e_d$),  
for different values of $\e_d$ ($-0.005 \le \e_d \le -0.0001$). Fig.~\ref{fig2}(d) 
shows a zoom of the results in panel (c) close to the lowest values of $\mu$. 
Notice that the red squares curve at the top for $\e_d=-0.005$ is at the PHS point for $\mu=0$, 
and the occupancy is pinned at $n_d=1$ even as $\mu$ moves away from PHS. 
As expected, the average value of $n_d$ decreases (from $n_d \approx 1.0$ to $n_d \approx 0.45$) 
as $\e_d$ increases, from $\e_d=\nicefrac{-U}{2}=-0.005$ (red squares) 
to very close to the Fermi energy$, \e_d=-0.0001$ (light green left-triangles), 
moving the system from deep into the Kondo regime to an intermediate valence regime, even for $\mu=0$.  
However, as the Fermi energy approaches the bottom of the band ($\mu \approx -1.0 - \e_d$), 
$n_d$ increases abruptly, with a faster rate the closer $\e_d$ is to zero. 

The variation of the bound state spectral weight $Z_b$~\cite{zb} 
[purple hexagons curve in Fig.~\ref{fig2}(c)] with $\mu$, for the non-interacting case, 
indicates that once the system moves to the intermediate valence regime 
(larger values of $\e_d$), the presence of the bound state below the bottom of the band, 
with stronger spectral weight, strongly increases the charging of the impurity. 
This effect is negligible if the system is well into the Kondo regime 
($\e_d=-0.005$, red squares) or close to it ($\e_d \lesssim -0.003$). 

\begin{figure}[h]
\includegraphics[width=1.0\columnwidth]{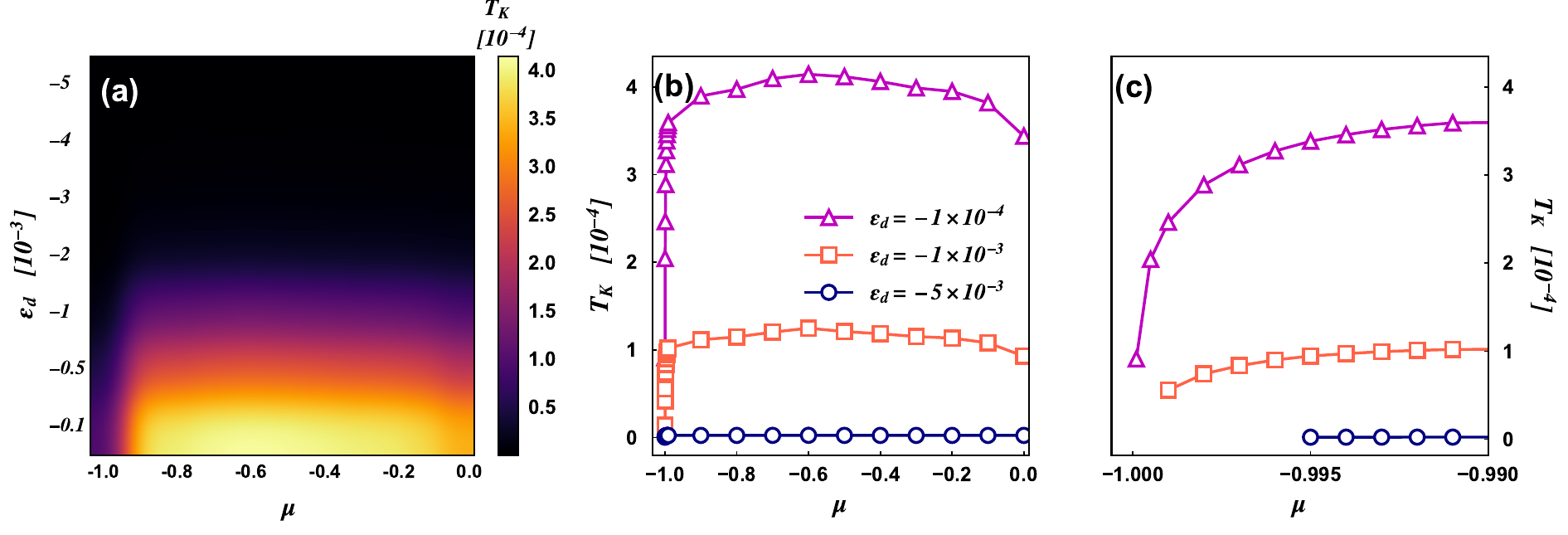}
	\caption{\label{fig3}(a) Color map of the Kondo 
	temperature $T_K$ for all the points in Fig.~\ref{fig2}(c).  
	The sudden drop of $T_K$ at the bottom 
	left corner indicates that the bound state near $E_F$ 
	moves the system back to the Kondo regime, which has a lower 
	$T_K$ than in the intermediate valence regime (bottom right corner). 
	(b) Comparison of $T_K$ vs $\mu$ results 
	for $\e_d=-1 \times 10^{-4}$ (magenta triangles, intermediate valence 
	regime), $\e_d=-1 \times 10^{-3}$ (orange squares, border between 
	Kondo and intermediate valence regimes), and $\e_d=-5 \times 10^{-3}$ 
	(blue circles, deep into Kondo regime). (c) Zoom of the results in 
	panel (b) close to the bottom of the band. 
	} 
\end{figure}

As $n_d$ tends to approach its 
Kondo-value of $n_d \approx 1.0$, despite $\e_d$ approaching the intermediate valence regime, close to 
the bottom of the band, it is reasonable to expect that the Kondo 
temperature $T_K$ will be strongly affected. We expect $T_K$ will tend to return to its 
Kondo-value when we approach the bottom of the band in the 
intermediate valence regime. Indeed, this can be seen in Fig.~\ref{fig3}(a), 
showing a color map of the Kondo temperature $T_K$ for all the points 
in Fig.~\ref{fig2}(c). Focusing on the right side of the figure ($\mu=0$, $E_F$ 
at the center of the band), we see 
the usual increase in $T_K$ as we move from top to bottom (from the Kondo 
to the intermediate valence regime). However, looking at the bottom 
of the figure, moving from right to left (from center to 
bottom of the band) we see that $T_K$ decreases abruptly as 
we approach the bottom of the band, tending back to its low Kondo-regime  
value. Figure \ref{fig3}(b) shows a comparison of results for $T_K$ vs $\mu$ 
for different $\e_d=-1 \times 10^{-4}$ (magenta triangles, intermediate valence regime), 
$\e_d=-1 \times 10^{-3}$ (orange squares, border between Kondo and intermediate valence regimes), 
and $\e_d=-5 \times 10^{-3}$ (blue circles, Kondo regime), highlighting the sharp drop in $T_K$ 
as $\mu$ approaches the singularity, when the system is in the intermediate valence regime (magenta triangles) 
or in a region in between Kondo and intermediate valence~\cite{Zilatic2010} (orange squares). 
This contrasts the stable behavior of $T_K$ when the system is 
deep into the Kondo regime (blue circles). Figure \ref{fig3}(c) shows a 
zoom of the results close to the singularity. Indeed, the formation of the bound state 
close to the Fermi energy seems to bring the system back to a Kondo regime~\cite{kondotemp}. 

\subsection{Evolution of the bound state with correlations}\label{sec-correl}

\begin{figure}[h]
\includegraphics[width=1.0\columnwidth]{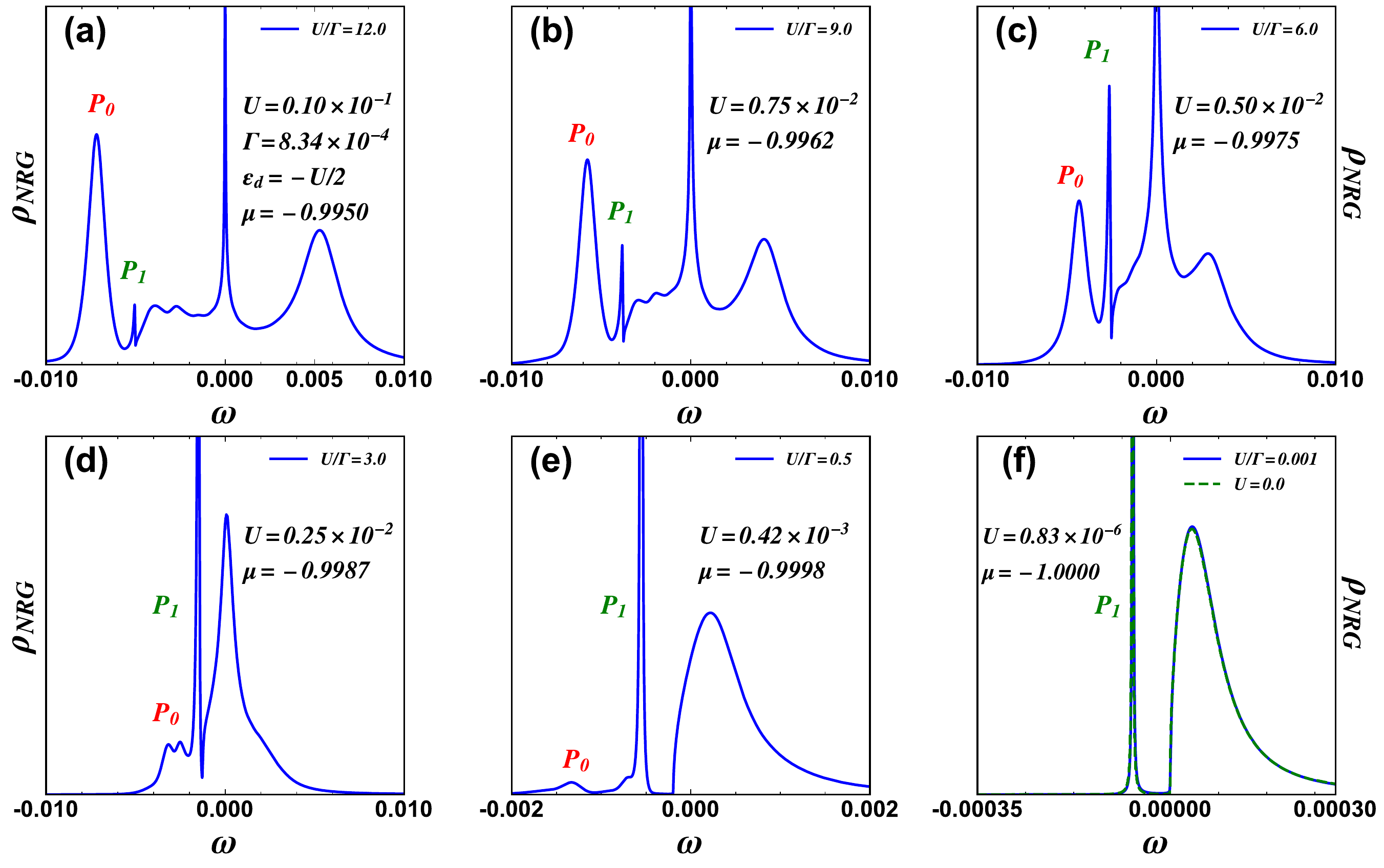}
	\caption{\label{fig4} Evolution of the interacting impurity spectral density $\rho_{NRG}$
	(blue curves), for $\Gamma=8.34 \times 10^{-4}$, 
	$\e_d=\nicefrac{-U}{2}$, and $\mu=-1.0 -\e_d$, as $U$ varies from 
	$\nicefrac{U}{\Gamma}=12$ [panel (a)] to $0.001$ [panel (f)]. 
	The value of $\mu$ places $\e_d$ at the singularity in all panels. 
	The dashed green curve in panel (f) 
	is the non-interacting $\rho_{U=0}$ spectral function, showing 
	excellent agreement with the $U \approx 0$ NRG result. The peaks $P_0$ 
	and $P_1$ are associated with the bound state seen in $\rho_{U=0}$ (see text). 
	} 
\end{figure}

With the objective of understanding the origin of the split peaks 
around the singularity visible in Fig.~\ref{fig2}(b), 
we present in Fig.~\ref{fig4} the evolution of the 
interacting impurity spectral function as $U$ decreases, 
keeping $\Gamma = 8.334 \times 10^{-4}$, $\e_d=\nicefrac{-U}{2}$, and 
varying $\mu$ so that, for all panels, $\e_d$ is at the singularity ($\mu=-1-\e_d$). Panels 
(a) to (f) show results for $\nicefrac{U}{\Gamma} =12.0$, 
$9.0$, $6.0$, $3.0$, $0.5$, and $0.001$, respectively ( as 
indicated in each panel). With decreasing $U$, the upper CBP moves to lower energy, 
eventually merging with a considerably broader Kondo peak, [panel (c)],
resulting from the system having entered an intermediate valence regime~\cite{kondotemp}.
We now focus our attention on the two peaks below the bottom 
of the band, whose position and spectral weight can be followed 
more accurately [peaks $P_0$ and $P_1$]. 
For decreasing $\nicefrac{U}{\Gamma}$, the 
leftmost peak, $P_0$, transfers its spectral 
weight to peak $P_1$, located at the bottom of the band. 
Indeed, for $\nicefrac{U}{\Gamma} \lesssim 0.5$ [panel (e)],  
$P_0$ has transferred almost all of its spectral weight to $P_1$,  
while in the interval $3 \lesssim \nicefrac{U}{\Gamma} \lesssim 6$, 
peak $P_0$ splits into two peaks.
For $\nicefrac{U}{\Gamma} \approx 9.0$, $P_1$ detaches from the 
bottom of the band, moving away from it for smaller $U$, while 
its spectral weight increases at the expense of $P_0$. Panel (f), 
for $\nicefrac{U}{\Gamma}=0.001$, 
has a comparison of the NRG (blue curve) and $U=0$ results~\cite{dirac} 
(dashed green curve), showing that they are virtually the same. This demonstrates 
that the NRG spectral function results reproduce faithfully 
the evolution of the many-body processes that give origin to the 
split peaks around the singularity, deep into the Kondo regime 
(panel (a), $\nicefrac{U}{\Gamma}=12$).

We now analyze $\rho_{NRG}$ in more detail in panels (a) and (b),  
where we still have strong correlations. In both panels,  
a well formed Kondo peak and an upper CBP are clearly visible. For $\nicefrac{U}{\Gamma}=9$, 
at energies below the Kondo peak (but still inside the continuum), a 
structure with two features is clearly visible. For $\nicefrac{U}{\Gamma}=12$, 
the Kondo peak and the upper CBP are clearly consolidated and further 
structure (a third smooth feature) emerges between the Kondo peak and the 
bottom of the band. In Appendix~\ref{appD}, we will show that all these 
features (including $P_0$ and $P_1$) are not NRG numerical artifacts.

\begin{figure}[h]
\includegraphics[width=1.0\columnwidth]{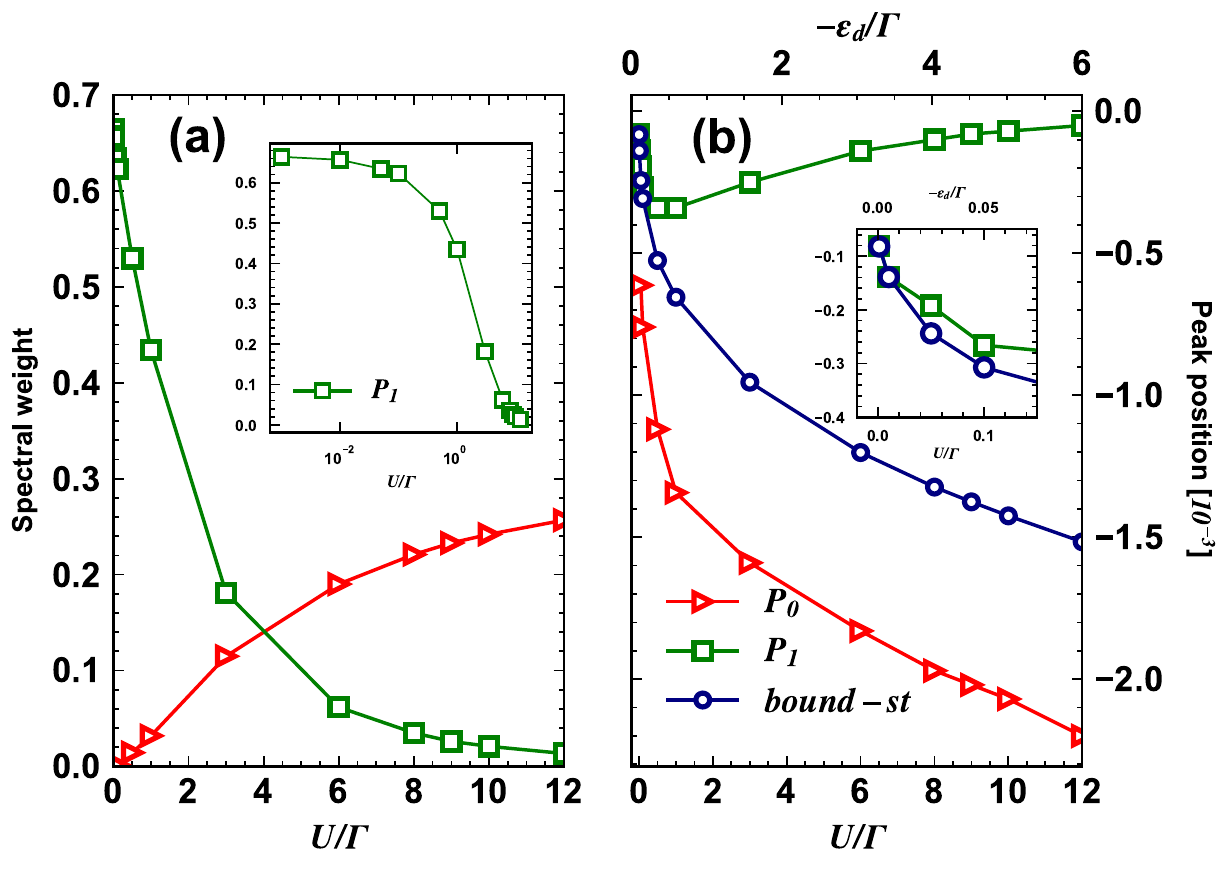}
	\caption{\label{fig5} (a) Spectral weight, as a function of $\nicefrac{U}{\Gamma}$, 
	for peak $P_0$ (red right-triangles) and peak $P_1$ (green squares), as defined in the 
	discussion of Fig.~\ref{fig4}. The inset shows $P_1$ spectral weight in a 
	$\log$ scale, highlighting the $\nicefrac{2}{3}$ plateau for vanishing $U$. 
	(b) Energy position, in relation to the bottom of 
	the band, of $P_0$ (red right-triangles), $P_1$ (green squares), and 
	the non-interacting bound-state (blue circles). The inset shows the 
	exact agreement between $P_1$ and the bound-state positions for the 
	smallest values of $U$. Note that the non-interacting ($U=0$) bound-state results 
	(blue circles) are dependent on $\nicefrac{-\e_d}{\Gamma}$ (upper horizontal axes 
	in the main panel and its inset), with $\e_d=\nicefrac{-U}{2}$, 
	where the $U$ and $V$ values are defined by the NRG results. 
	} 
\end{figure}

Fig.~\ref{fig5} presents the spectral weight [panel (a)] and position [panel (b)], 
in relation to the bottom of the band, of peaks $P_0$ and $P_1$ for different values in the interval 
$0.001 \le \nicefrac{U}{\Gamma} \le 12.0$. In panel (b), we also plot the position 
of the non-interacting bound-state (blue circles), for $U=0$ and same $\e_d$ and $V$ values used in the 
NRG calculations (note that the non-interacting results depend on $\nicefrac{-\e_d}{\Gamma}$, which 
labels the upper horizontal axes in panel (b) and its inset). 
Following the spectral weight curves for peaks $P_0$ (green squares) and 
$P_1$ (red right-triangles), in Fig.~\ref{fig5}(a), we see that, when $\nicefrac{U}{\Gamma}$ 
decreases from $12$ to $0.5$, the spectral weight of 
$P_0$ is almost all transferred to $P_1$ (although part of $P_0$'s spectral 
weight is also transferred to the continuum, and then, with 
further decrease of $\nicefrac{U}{\Gamma}$, to peak $P_1$). 
In the interval $0.5 \ge \nicefrac{U}{\Gamma} \ge 0.001$, 
$P_1$ quickly acquires spectral weight from inside 
the continuum, reaching $\approx \nicefrac{2}{3}$ for very 
small values of $\nicefrac{U}{\Gamma}$, as expected. The inset in Fig.~\ref{fig5}(a), 
shows $P_1$'s spectral weight in a $\log$ scale to emphasize the 
formation of a $\nicefrac{2}{3}$ plateau as $U \rightarrow 0$.

\begin{figure}[h]
\includegraphics[width=1.0\columnwidth]{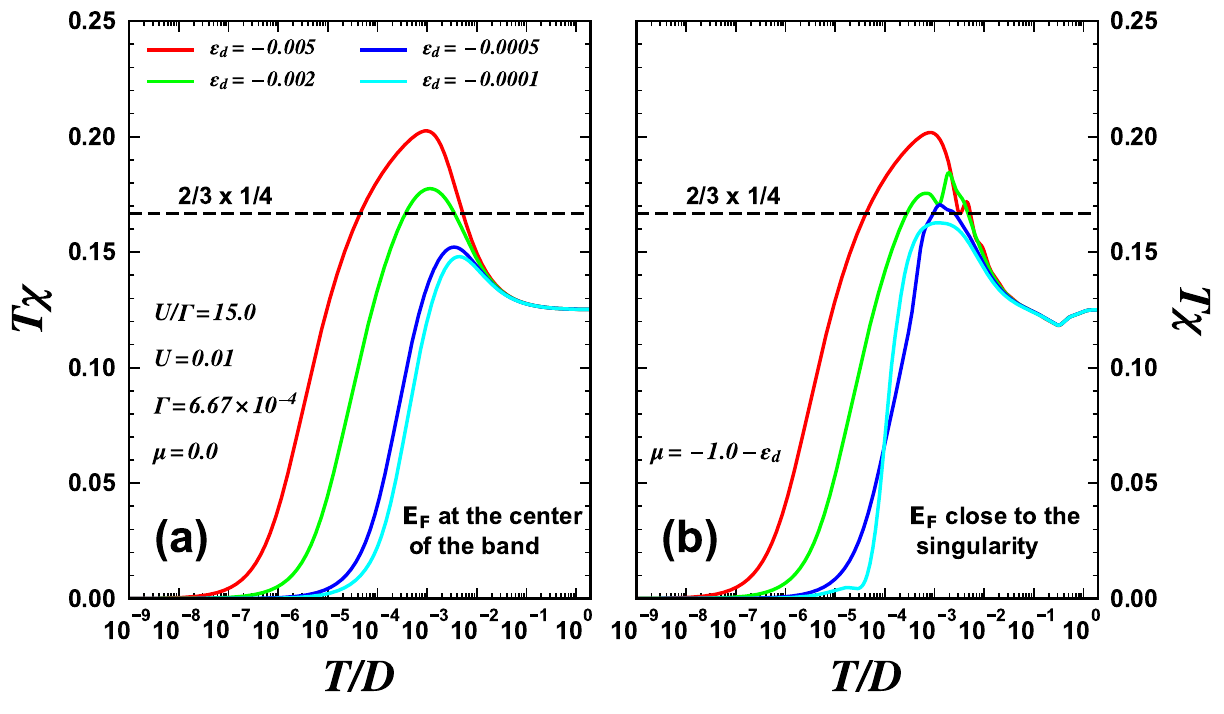}
	\caption{\label{fig6} Impurity magnetic susceptibility $\chi$
	(as a function of temperature, in units of $D$, half bandwidth), for some of the $\e_d$ values 
	in Fig.~\ref{fig2} (d). In panel (a), the Fermi energy is fixed at the center of the band and, 
	as $\e_d$ gradually increases, the system moves into the intermediate valence regime. 
	Note that for $\e_d=-0.005$, the system is in the PHS point. (b) Same as in (a), but now 
	the chemical potential is such that $\e_d$ remains fixed at the singularity and it is the 
	Fermi energy that moves closer to the bottom of the band, thus $\mu=-1.0 - \e_d$. The 
	values of $\Gamma$ and $U$ are the same as in Fig.~\ref{fig2} for all calculations.
	} 
\end{figure}

Figure \ref{fig5}(b) shows the evolution of the position of 
$P_0$ (red right-triangle) and $P_1$ (green squares), 
measured in relation to the bottom of the band. Their 
variation in position is contrasted to that of the non-interacting bound-state (blue circles).  
Starting from $\nicefrac{U}{\Gamma}=12$, $P_1$ (green squares) moves away from the bottom of 
the band as $\nicefrac{U}{\Gamma}$ decreases, until, at $\nicefrac{U}{\Gamma} \approx 1$, 
it reverses course and starts to approach the bottom of the band again. The position 
of $P_1$, the dominant peak for small values of $U$, progressively approaches the position of 
the bound-state, until they coincide for the two smallest values of $U$, 
as emphasized in the inset. Peak $P_0$, on the other hand, monotonically approaches the bottom 
of the band as $\nicefrac{U}{\Gamma}$ decreases, initially linearly, but then, 
around the same region where $P_1$ reverses course, starts to show a 
faster rate of approach to the bottom of the band as a function of 
$\nicefrac{U}{\Gamma}$. Note that, to simplify the presentation, 
even after $P_0$ split into two peaks, we are considering it as a single 
peak and taking a point halfway between the split peaks as $P_0$'s position.

Finally, the results for the position of $P_1$ [red right-triangles in panel (b)] 
can be interpreted in the following way. As $U$ decreases at fixed $\Gamma$, the Fermi energy approaches the 
bottom of the band, since $\epsilon_d=\nicefrac{-U}{2}$ is at the 
singularity. Since the Kondo peak becomes broader (as $\nicefrac{U}{\Gamma}$ 
decreases), $P_1$ is initially slowly forced away from the bottom of the band. 
However, for very small $U$ values ($\nicefrac{U}{\Gamma} \lesssim 1$), 
the Fermi energy gets very close to the singularity, 
and, since $\Gamma = \pi \rho_0 V^2$ is fixed, $V$ decreases (since $\rho_0$ 
increases), and $P_1$, which has become the bound-state (check comparison with blue circles 
curve in the inset), approaches the bottom of the band again (check also 
$\epsilon_b$ in Fig.~\ref{fig15}).

\subsection{Thermodynamic properties: the fractional local moment}\label{sec-thermo}

Figure \ref{fig6} shows the impurity magnetic susceptibility for 
four values of $\e_d$ ($-0.005$, $-0.002$, $-0.0005$, and $-0.0001$).  
In panel (a), the Fermi energy is fixed at the center of the band ($\mu = 0$) 
and the system stays in a more standard Kondo regime. In contrast, 
in panel (b), $\mu = -1.0 - \e_d$, $E_F$ is located close to the bottom of the band, 
such that $\e_d$ is at the singularity for all cases. Starting at the PHS point 
$\e_d = -0.005$ [red curve, panel (a)], the system progressively moves into 
the intermediate valence regime as $\e_d$ increases ($\e_d$ 
approaches the Fermi level). As expected, the broad peak around 
$T \approx 0.001 << U=0.01$, seen in the red curve, is 
indicative of the LM fixed point. 
As $\e_d$ moves closer to the Fermi energy, charge fluctuations 
become more prominent, suppressing the formation of a LM 
at the impurity (cyan curve). On panel (b), however, 
where $\e_d$ is at the singularity, and the Fermi energy progressively 
approaches it, the picture that emerges is substantially 
different. Although there is little difference between both panels 
for $\e_d=-0.005$ (red curves), once the Fermi energy approaches 
$\e_d$, the suppression of the LM peak seems to be arrested in panel (b). The LM peak in $\chi$ stays 
pinned close to the $\nicefrac{2}{3} \times \nicefrac{1}{4}$ value, 
indicative of the presence of the bound state with spectral 
weight $Z_b=\nicefrac{2}{3}$ (see Appendix), even as $\e_d$ 
changes by an order of magnitude. 

\begin{figure*}[!htbp]
\includegraphics[width=1.9\columnwidth]{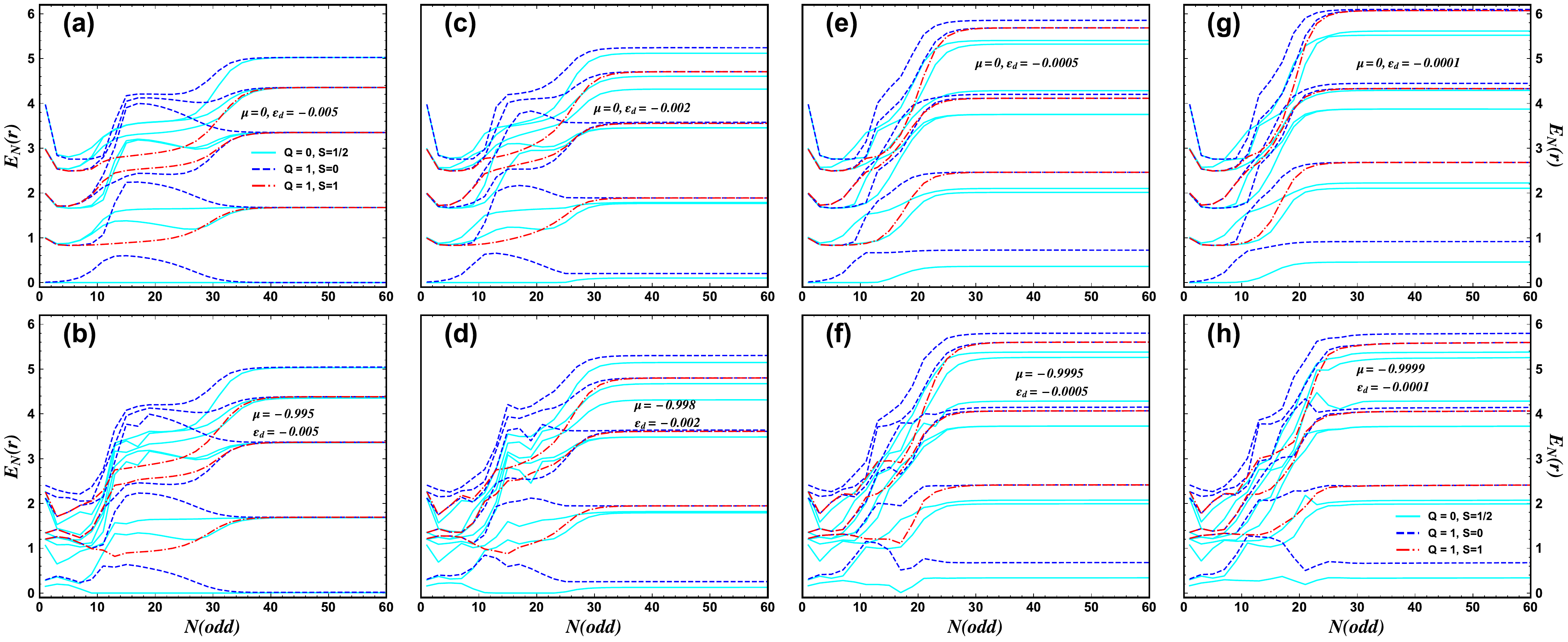}
	\caption{\label{fig7} NRG energy flow (for odd values of N) 
	for the same parameters as in Fig.~\ref{fig6}. The $\e_d$ values, 
	from left to right, are: $-0.005$, 
	$-0.002$, $-0.0005$, and $-0.0001$. Top panels 
	correspond to the Fermi energy at the center of the band; 
	lower panels correspond to a Fermi energy close to the bottom 
	of the band (with $\e_d$ at the singularity). Q is the  
	charge in the system, measured in relation to half-filling, 
	while $S$ indicates the total spin (see Ref.~\cite{Bulla2008} 
	for details). The parameter values for $U$ and $\Gamma$ are 
	the same as in Fig.~\ref{fig6}. For these NRG calculations, 
	in particular, we have used $\Lambda=2.5$.
	} 
\end{figure*}

The behavior of $\chi$ shown in Fig.~\ref{fig6}(b) suggests that 
the existence of the bound state makes the LM impervious 
to charge fluctuations. This is corroborated by the NRG 
energy flows, as shown in Fig.~\ref{fig7}, for the same parameters as in Fig.~\ref{fig6}. 
It is well-known that the three SIAM fixed points 
are associated to energy plateaus in the spectra as the number  
of NRG iterations varies. The Free Orbital, Local Moment, and Strong 
Coupling fixed points are successively approached as $N$ increases (which corresponds 
to a decrease in temperature, or energy scale behavior). This can be easily spotted in Fig.~\ref{fig7}(a), 
corresponding to the PHS point, see Refs.~\cite{Krishna-murthy1980,Bulla2008} 
for comparison. In the upper-row panels (Fermi energy at the 
center of the band, $\mu=0$), we see that the plateaus starting at approximately 
$N=10$ [Fig.~\ref{fig7}(b)] are gradually erased as charge fluctuations increase 
[panels (c), (e), and (g)]. For example, in Fig.~\ref{fig7}(g), the lowest energy state with 
$Q=1$ and $S=0$ (dashed blue curve) transitions directly (around $N=10$) 
from a high-temperature value to a low-temperature value without going through 
an intermediate stage. This does not happen for the lower-row panels ($E_F$ 
close to the bottom of the band). 
The plateau present in Fig.~\ref{fig7}(b) (between $N=10$ and $N=30$) 
is still present in  panel (h) (although it now finishes at 
around $N=20$). This is consistent with the results for a robust Kondo state 
seen in the impurity magnetic susceptibility in Fig.~\ref{fig6}. 

\begin{figure}[h]
\includegraphics[width=1.0\columnwidth]{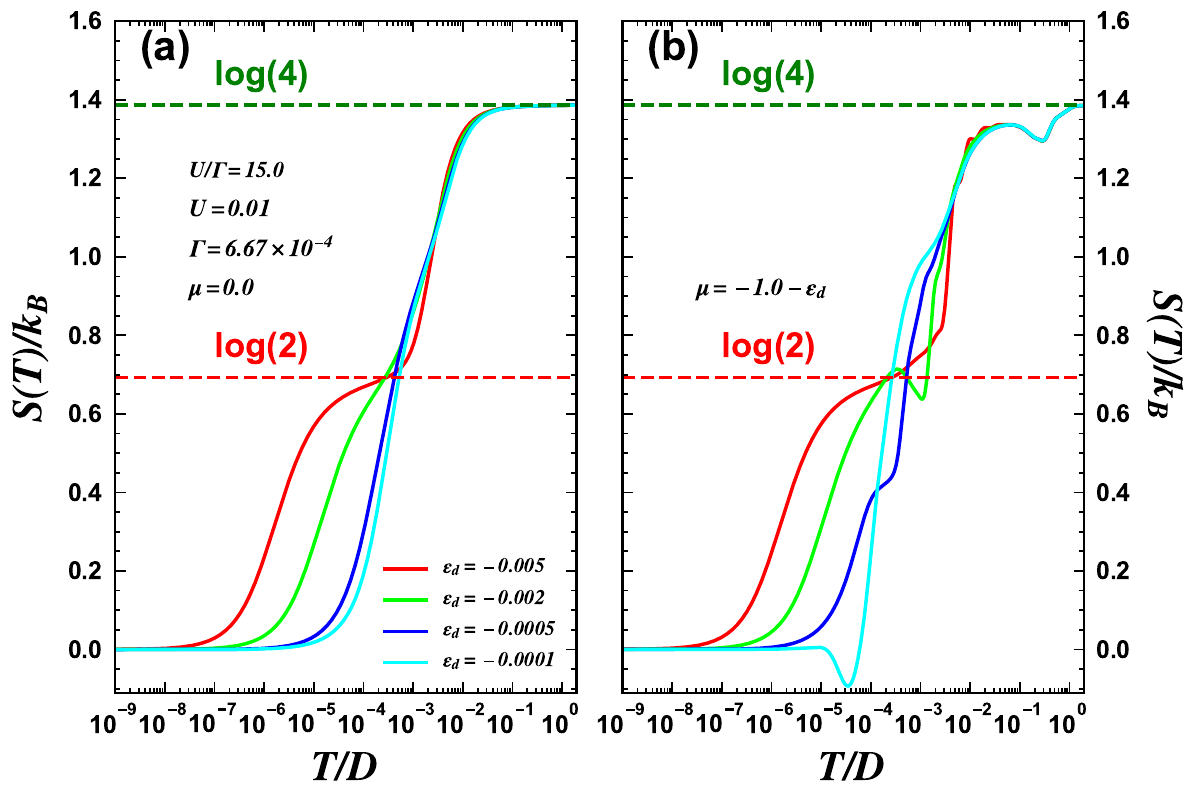}
	\caption{\label{fig8} Impurity entropy 
	(as a function of temperature) for the same parameters as  
	in Fig.~\ref{fig6}. See text for discussion of the negative 
	values in the cyan curve in panel (b).  
	} 
\end{figure}

For completeness, Fig.~\ref{fig8} shows the impurity 
entropy as a function of $\nicefrac{T}{D}$, for the same parameters 
as in Fig.~\ref{fig6}. We note in panel (a) ($E_F$ at the center 
of the band), for $\e_d=-0.005$ (red curve), the usual evolution. 
As temperature decreases, the impurity entropy goes from the Free Orbital 
($\ln 4$) plateau, to the Local Moment ($\ln 2$) plateau, until it reaches 
the Strong Coupling ($\ln 1$) Kondo limit. Panel (b) shows the corresponding 
results when the Fermi energy is close to 
the bottom of the band, with $\e_d$ at the singularity, as in Fig.~\ref{fig6}. 
We notice again a pinning tendency around the Local Moment plateau as $\e_d$ 
increases, specially for $\e_d=-0.002$ (green curve), whose oscillation, also 
observed in the corresponding result in Fig.~\ref{fig6}, may be ascribed to the 
many-body states that appear between the Kondo peak and the singularity 
[see Fig.~\ref{fig2}(b)]. It is important to note the non-universal low-temperature behavior 
of the cyan curve in panel 6(b), which shows the impurity entropy 
being negative in the range $10^{-5} \lesssim \nicefrac{T}{D} \lesssim 10^{-4}$.  
This behavior (which accompanies the non-monotonic behavior of the 
cyan curve in $T\chi$, Fig.~\ref{fig6}), is reminiscent of the 
behavior seen in other Kondo problems in the presence of a sharp 
singularity or discontinuity in the DOS~\cite{Zhuravlev2009,Zhuravlev2011, 
Mastrogiuseppe2014}, and is clearly most prominent here when the RL 
is closest to the band singularity.

\section{Results for an armchair graphene nanoribbon}\label{sec-nanoribbon}

\begin{figure}[h]
\includegraphics[width=1.0\columnwidth]{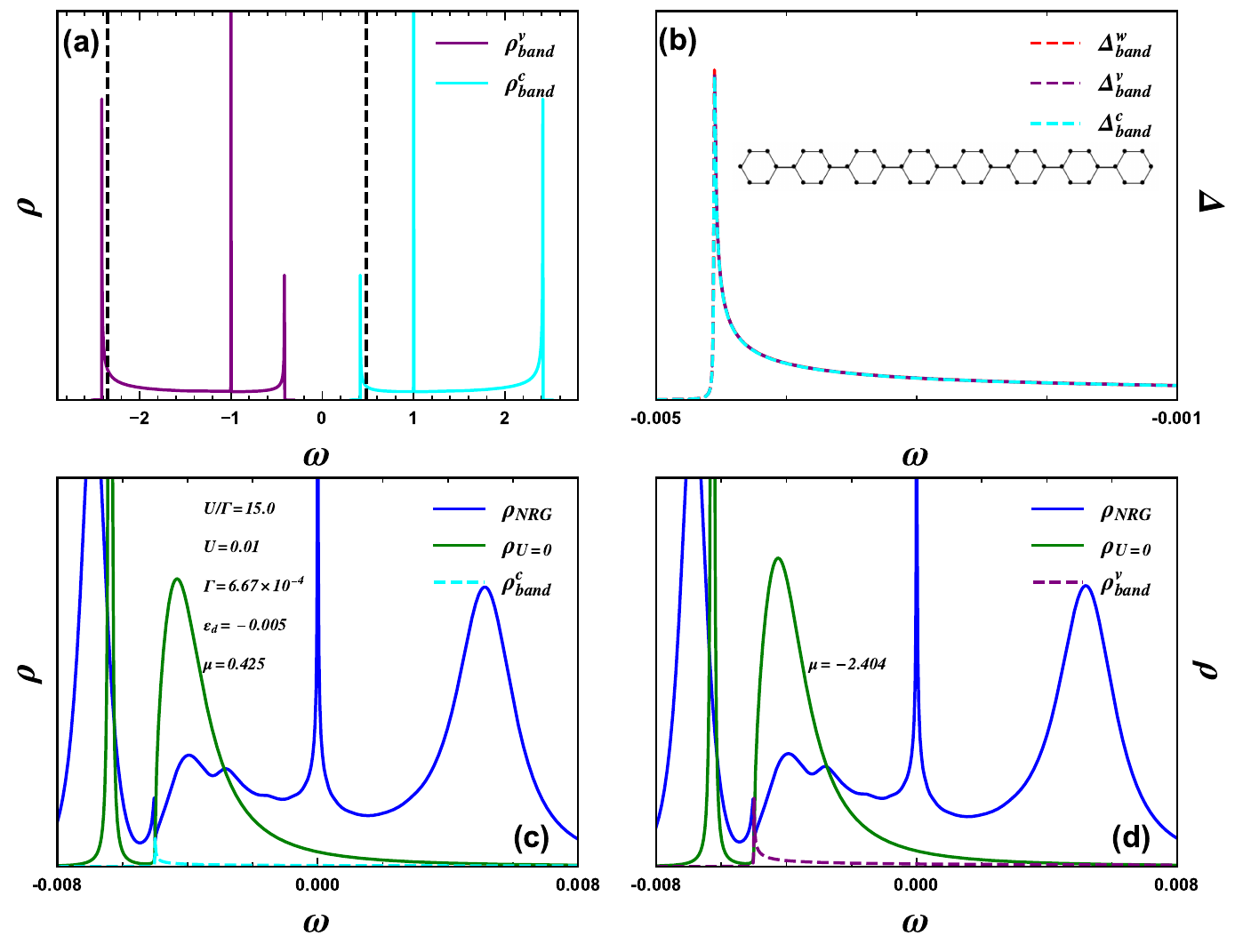}
	\caption{\label{fig9} (a) 3-AGNR DOS, showing the very symmetric valence (purple curve) 
	and conduction (cyan) bands. 
	The vertical black dashed lines indicate the approximate position of 
	the Fermi energy for the calculations in panels (c) and (d). (b) Comparison of the quantum wire 
	singularity ($\Delta_{band}^w$, dashed red curve) with the 3-AGNR valence 
	($\Delta_{band}^v$, dashed purple) and conduction ($\Delta_{band}^c$, dashed cyan) 
	hybridization functions lower singularities, showing they are identical. 
	The inset shows a sketch of the 3-AGNR system. 
	(c) Impurity DOS (blue curve, $\rho_{NRG}$) when $\e_d$ is at the singularity 
	at the bottom of the conduction band. The result is very similar to 
	Fig.~\ref{fig2}(b), for the quantum wire singularity. (d) Same as in (c), but now $\e_d$ is 
	at the valence band singularity. Also similar to panel (c). Parameter values (except for $\mu$) 
	are as in Fig.~\ref{fig2}(b). 
	} 
\end{figure}

We now discuss a possible physical implementation of the model 
NRG results presented in the previous sections. We 
study the Kondo states associated to band-edge singularities present in 
the DOS of an armchair graphene nanoribbon made by three carbon rows (3-AGNR), which is 
known to be a semiconductor~\cite{Lin2009,Almeida2022}. Such nanoribbons 
can be fabricated from molecular precursors~\cite{Cai2010}, for example, and have been used 
to study Kondo resonances in experiments~\cite{Li2017}. One interesting recent study is 
that of subgap states in a Kondo regime~\cite{Zalom2022}. 
Figure \ref{fig9}(a) shows the DOS of an undoped 3-AGNR with symmetric valence 
and conduction bands. The Fermi energy can in principle be gated down until it is close to the 
bottom singularity of the valence band (leftmost vertical dashed line), or we may gate-dope it 
with slightly more electrons and bring the Fermi energy just above the bottom singularity 
of the conduction band (rightmost vertical dashed line). Both cases reproduce the situation studied in the 
previous sections, for the quantum wire. Figure \ref{fig9}(b) compares the $\omega$-dependence of 
these two singularities with that of the quantum wire, showing that 
they are virtually the same. Notice that panel (b) shows the 
hybridization function $\Delta(\omega)=\pi V^2 \rho(\omega)$, such that 
$\Gamma=\pi V^2 \rho(0)$ is the same for all three cases~\cite{5-AGNR}. 
These results in Fig.~\ref{fig9}(b) imply that the NRG results for 
the 3-AGNR and the quantum wire should be very similar. 
Indeed, the finite-$U$ impurity spectral function [same parameters as in Fig.~\ref{fig2}(b)], 
shown in Fig.~\ref{fig9}(c), for the conduction band singularity, is 
quantitatively similar to the quantum wire results in Fig.~\ref{fig2}(b). The same occurs 
for the valence band singularity [Fig.~\ref{fig9}(d)]~\cite{flat}. 

\section{Summary, Discussion, and conclusions}\label{sec-conc}
We have analyzed the effect of Van Hove singularities 
near the Fermi energy and a magnetic impurity on the Kondo effect. Such singularities are 
present at the band edges of a quantum wire and of different AGNRs. 
The singularities present at the bottom of the valence and conduction bands of a 3-AGNR  
result in effective hybridization functions (at fixed $\Gamma$) 
with an $\omega^{\nicefrac{-1}{2}}$ dependence. Thus the spectral functions of the 
magnetic impurity are quantitatively similar to those obtained for a quantum wire 
and provide a convenient physical implementation of our model 
calculations~\cite{Kitao2020,Cai2010,Li2017}. 
The main results we obtained are as follows. For a non-interacting impurity, we have 
characterized a Dirac delta bound state below the band 
minimum, with properties that depend on the near-resonance of $\e_d$ with 
the singularity, and on the coupling of the impurity to the band. 
As expected, the larger the coupling and the closer $\e_d$ is to the singularity, the larger is the 
spectral weight $Z_b$ of the bound state and the farther it is below the 
band-minimum. The spectral weight $Z_b$ is vanishingly small if $\e_d$ is not 
close to the singularity, while it quickly increases as it approaches, reaching the 
value $Z_b=\nicefrac{2}{3}$ at resonance, in agreement with previous work~\cite{Agarwala2016}, 
where the singularity is due to spin-orbit interaction. In addition, the impurity level $\e_d$ 
is slightly renormalized upwards due to its interaction with the singularity (see Figs.~\ref{fig15}, 
\ref{fig18}, and \ref{fig20} in the Appendix). 

Once the Hubbard $U$ is present, we see several interesting effects. 
First, starting with the Fermi energy at the PHS point 
in the middle of the band [Fig.~\ref{fig2}(a)] and then moving 
to the bottom of the band [Fig.~\ref{fig2}(b)], at fixed $U$ and $\Gamma$, 
we see that the non-interacting bound state acquires a finite width~\cite{finitewidth} 
and moves further away from the bottom of the band; in addition, a discontinuity 
appears in the impurity DOS at the band edge. Additional structure in the 
spectral function appears between this discontinuity and the 
Kondo peak, which, aside from acquiring some asymmetry, it is barely 
affected. An analysis of the evolution of the impurity DOS as 
$U$ decreases, at fixed $\Gamma$, from $\nicefrac{U}{\Gamma}=12$ 
to $10^{-3}$ (see Fig.~\ref{fig4}) allows us to follow the evolution 
of these many-body-related features, until the $\nicefrac{U}{\Gamma}=10^{-3}$ results match perfectly the 
$\nicefrac{U}{\Gamma}=0$ results. 

An analysis of how the impurity is discharged as $\e_d$ moves 
closer to the Fermi energy, starting at the PHS point, 
shows that there is a great difference in the results for 
$\e_d$ being far from the singularity or close to the singularity. 
We see in Fig.~\ref{fig2}(c) that the approach of $\e_d$ to the singularity 
`recharges' the impurity, with the effect being more dramatic 
as it moves into the intermediate valence regime. 
This unusual behavior is clearly associated to the existence of the 
bound state. The Kondo temperature $T_K$ [see Fig.~\ref{fig3}(a)], 
for the same set of parameters, suffers a sizeable decrease 
(if in the intermediate valence regime, with $E_F$ 
at the center of the band) when $E_F$ moves closer 
to the singularity at the bottom of the band. Both results, 
on the occupancy of the impurity and its $T_K$ value, show that 
the system partially recovers its strong coupling 
regime properties, i.e., higher occupancy and lower $T_K$, 
once the presence of the singularity is felt at the intermediate 
valence regime. This occurs because of the formation of the bound state. 

In addition, the magnetic susceptibility shows that, for Fermi energy 
near the bottom of the band and in the intermediate valence regime, 
the LM fixed point is more resilient, as the impurity suppresses charge 
fluctuations. Figure \ref{fig6}(b) shows that the LM 
plateau is somewhat restored around the $\nicefrac{2}{3} \times 
\nicefrac{1}{4}$ value, indicating the influence of the 
bound state. This evolution is corroborated by an analysis 
of the NRG energy flow, shown in Fig.~\ref{fig7}, as well as 
the impurity entropy (Fig.~\ref{fig8}). 

\begin{acknowledgments}
We thank A.~Agarwala and V.~B.~Shenoy for illuminating 
exchanges at the beginning of this work, and discussions 
with E.~Vernek, N.~Sandler, and G.~Diniz. PAA thanks the Brazilian funding 
agency CAPES for financial support.
MSF acknowledges financial support from the National Council 
for Scientific and Technological Development (CNPq), grant number 
311980/2021-0, and from the Foundation for Support of Research in 
the State of Rio de Janeiro (FAPERJ), processes number 210 355/2018 
and E-26/211.605/2021. SEU acknowledges support 
from the U.S. Department of Energy, Office 
of Basic Energy Sciences, Materials Science and Engineering Division. 
GBM acknowledges financial support from the Brazilian agency Conselho
Nacional de Desenvolvimento Cient\'ifico e Tecnol\'ogico
(CNPq), processes 424711/2018-4, 305150/2017-0, and 210
355/2018.

\end{acknowledgments}

\appendix 

\section{Resonant level in the band continuum}\label{appA}

We wish to understand how the proximity of a RL (an impurity with $U=0$) to the singularity 
at the bottom of the 1D band affects the RL spectral function. 
We will show that the main effect of the singularity is the formation 
of a bound state out of the continuum, and then we will analyze 
its properties. 

We note that Ref.~\cite{Agarwala2016} discusses a bound state out of the continuum in 3D. 
There, however, it is \emph{necessary} to add spin-orbit interaction (SOI), while we show 
that this is not the case in 1D. Thus, we add SOI to the Hamiltonian 
presented in the main text, Eq.~\eqref{h1D}, by rewriting it as 

\begin{eqnarray}\label{Hwire}
	H_{\rm wire}&=\sum_k \Psi^\dagger_k{\cal H}_{\rm wire}(k) \Psi_k, 
\end{eqnarray}
where $\Psi^\dagger_k=(c^\dagger_{k\up}, c^\dagger_{k\dn})$, $c^\dagger_{k\sigma}$ creates 
an electron with wave vector $k$ and spin $\sigma=\up,\dn$, and ${\cal H}_{\rm wire}(k)$ is 
given by

\begin{eqnarray}
	{\cal H}_{\rm wire}(k) &=&  \left( -2t \cos k  - \mu \right)\sigma_0 + 
	\left(\beta \sigma_x + \alpha \sigma_y \right)2 \sin k , 
\end{eqnarray}
where $\beta$ and $\alpha$ are the 
Dresselhaus~\cite{Dresselhaus1955} and Rashba~\cite{Bychkov1984} SOI, 
respectively, while $\sigma_x$ and $\sigma_y$ are spin Pauli matrices and $\sigma_0$ 
is the $2 \times 2$ identity matrix.  It can be shown~\cite{Martins2020} that 
the energy dispersion associated to this Hamiltonian may be written as 
\begin{equation}
\e_{k\sigma}=-2\sqrt{t^{2}+\vert\gamma\vert ^{2}}\cos \left( k-\sigma\varphi\right) - \mu ,
\label{eq:eksigmar}
\end{equation}
where $\gamma=\beta + i\alpha$, $\varphi =\tan ^{-1}\left( \nicefrac{\vert\gamma\vert}{t}\right)$, 
and $\sigma = \pm$. The band structure and the DOS for this Hamiltonian are 
studied in the next section. 

\begin{figure}[h]
\includegraphics[width=1.0\columnwidth]{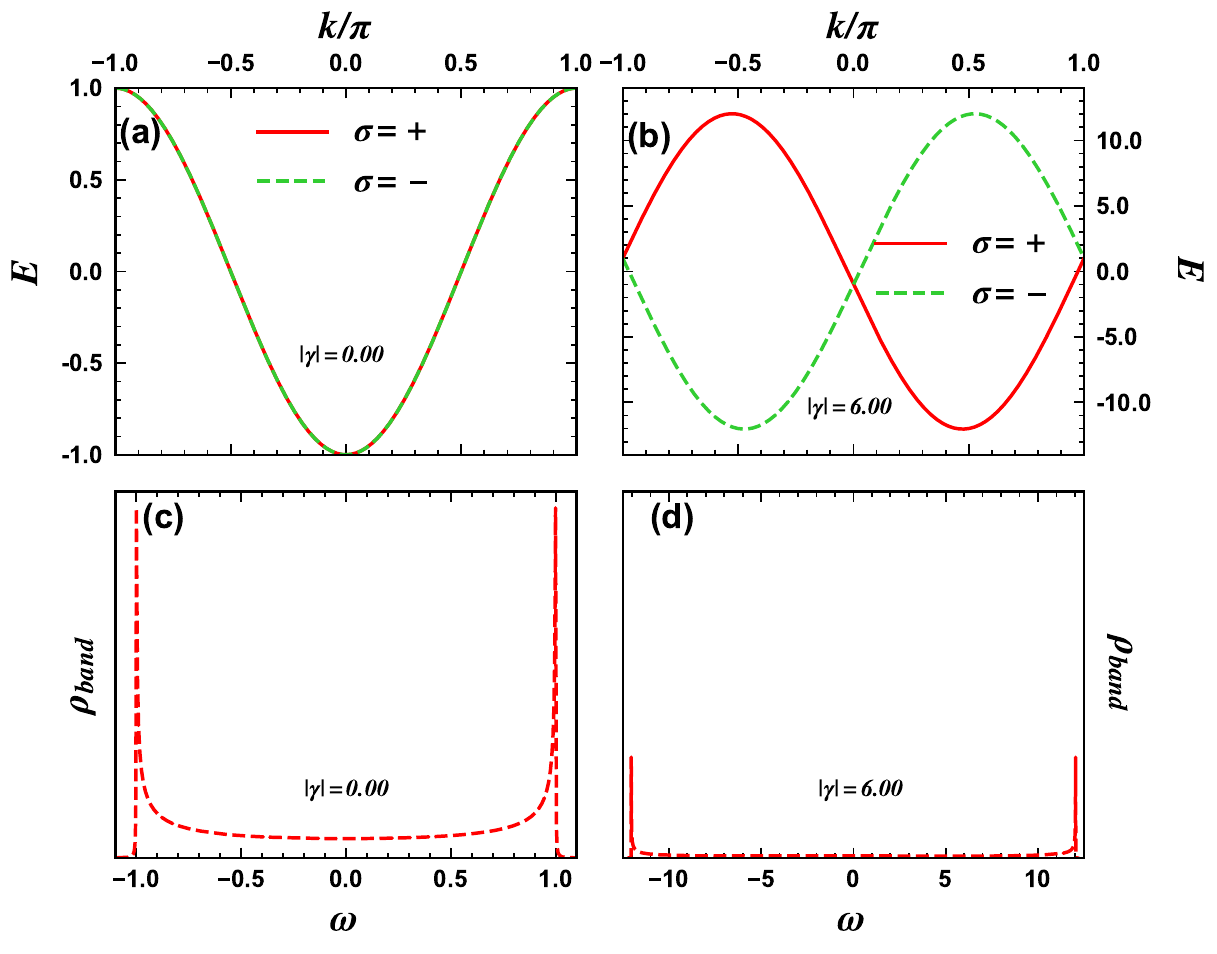}
	\caption{\label{fig10} Band structure for $\vert\gamma\vert=0.0$ and $\vert\gamma\vert=6.0$, 
	in panels (a) and (b), respectively. Corresponding band DOS for the same 
	values of $\vert\gamma\vert$, in panels (c) and (d). 
	Notice how the singularity for $\vert \gamma \vert=6.0$ 
	[panel (d)] carries a considerably smaller spectral weight than the corresponding singularity for 
	$\vert \gamma \vert=0.0$ [panel (c)]. Note that the range in the vertical axes in panels (c) and (d) 
	are the same, with the integral of $\rho_band$ equal to $1$ in both cases.
	} 
\end{figure}

\begin{figure}[h]
\includegraphics[width=1.0\columnwidth]{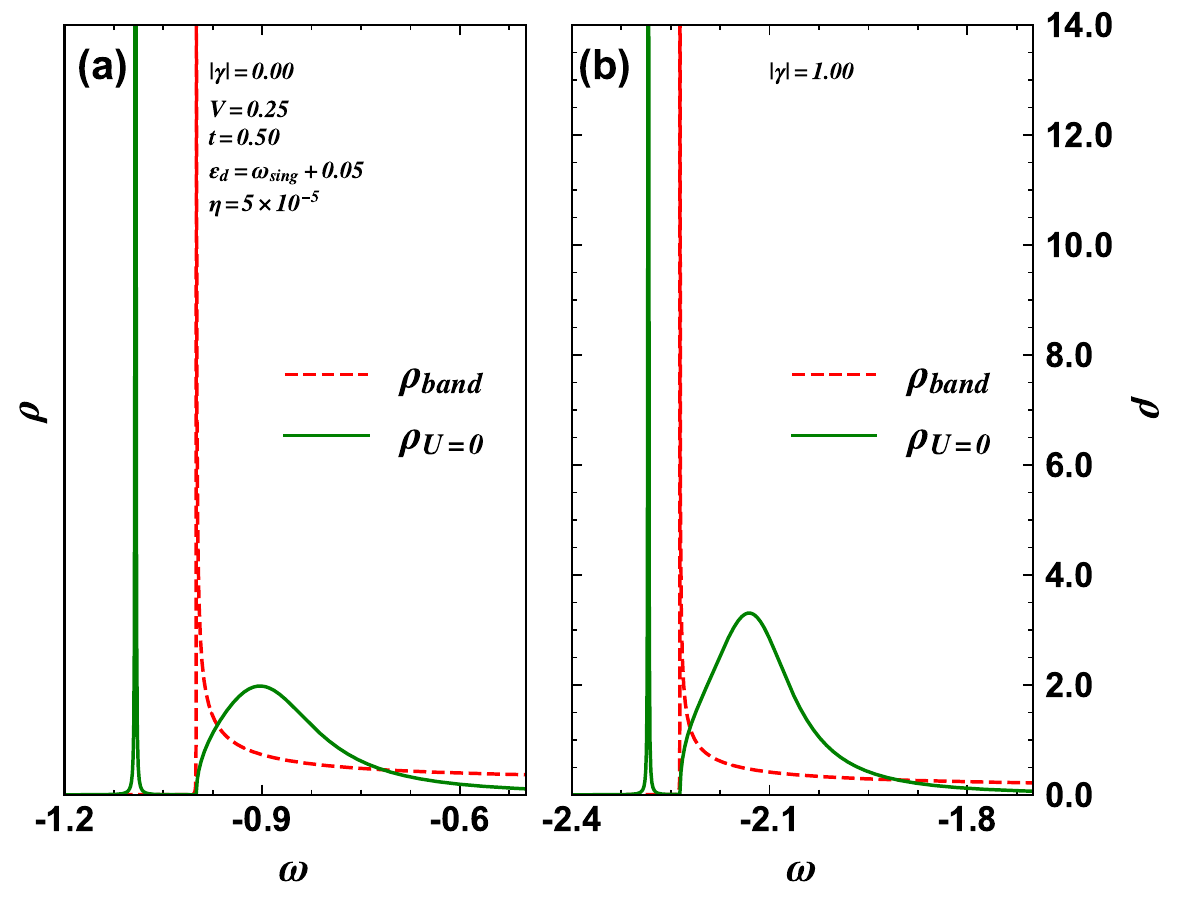}
	\caption{\label{fig11} Band DOS $\rho_{\rm band}$ (red dashed curves) and the RL  
	DOS $\rho_{U=0}$ (green curves) for $\vert\gamma\vert=0.0$ and $1.0$, in panels (a) and (b), respectively. 
	The RL orbital energy $\e_d=\omega_{sing}+\delta$ is 
	placed at $\delta=0.05$ above the bottom of the band $\omega_{sing}$, in both cases. In panel 
	(a), $Z_b=0.554$, while in panel (b), $Z_b=0.415$, showing that the increase of SOI 
	makes the bound state less bound and with a weaker spectral weight $Z_b$. 
	See text for a definition of $Z_b$.
	} 
\end{figure}

\section{Band structure and spectral function in 1D with 
spin orbit interaction}

In this section we show that the inclusion of SOI in 1D 
(which already has a singularity at the bottom of the band 
without SOI) decreases the spectral weight of the bound state. 
Figure \ref{fig10} shows the band structure 
in panels (a) ($\vert \gamma \vert=0.0$) and (b) 
($\vert \gamma \vert=6.0$), with the corresponding DOS in 
panels (c) ($\vert \gamma \vert=0.0$) and (d) 
($\vert \gamma \vert=6.0$). It is clear that SOI 
decreases the spectral weight carried by the singularity at 
the bottom of the band. This happens because a finite SOI 
increases the bandwidth [compare the range in the horizontal 
axes in Figs.~\ref{fig10}(c) and (d)]. 

\begin{figure}[h]
\center
\includegraphics[width=1.0\columnwidth]{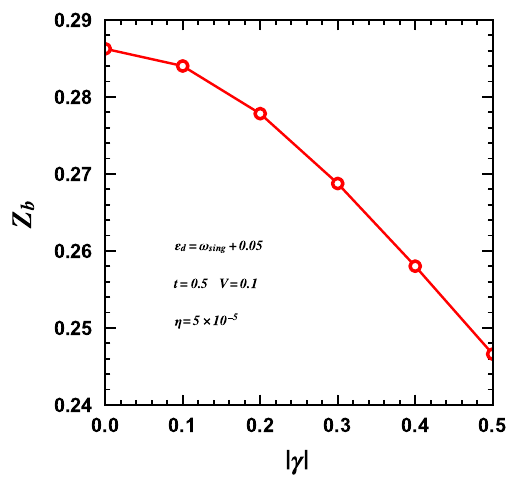}
	\caption{\label{fig12} Variation of the bound state spectral weight $Z_b$ 
	with $\vert\gamma\vert$ for $V=0.1$. 
 	} 
\end{figure}

We show next that this implies a 
loss of spectral weight of the bound state associated with the singularity. 

We calculate the RL Green's function $\hat G_{imp}(\omega)$, given by 

\begin{eqnarray}\label{Gimp}
	\hat G_{imp}(\omega)=\left[\left(\omega -\e_d\right)\sigma_0 -
	\hat{\Sigma}^{(0)}(\omega) + \mathrm{i} \eta \right]^{-1},
\end{eqnarray}
where $\hat{\Sigma}^{(0)}(\omega)=\sum_k \hat V \hat G_{wire}(k,\omega) \hat V^\+ $ 
is the hybridization self-energy,
with $\hat V=V\sigma_0$, $\hat G_{wire}(k,\omega)=\left[\omega \sigma_0-{\cal H}_{wire}(k) \right]^{-1}$ 
is the quantum wire Green's function, while $V$ is defined right after Eq.~(3) in the main text.  
The RL spectral function, i.e., its DOS, is calculated through (notation as in the main text)
$\rho_{U=0}(\omega)=-\frac{1}{\pi} \Im \mathrm{Tr}\, \hat{G}_{imp}(\omega)$.

A comparison of both the 1D lattice DOS (red curves) and the RL DOS (green curves) is 
shown in Fig.~\ref{fig11} for $\vert\gamma\vert = 0.0$ and $\vert\gamma\vert = 1.0$, panels (a) 
and (b), respectively, where we have set $\e_d=\omega_{sing}+\delta$ (where $\delta=0.05$ and 
$\omega_{sing}$ is the energy at the bottom of the band, with the band being 
symmetric around $\omega=0.0$), and $V=0.25$. A comparison of the 
RL DOS in both panels shows that a finite SOI decreases the 
spectral weight of the bound state [thus the area of the DOS inside the 
continuum in panel (b) is clearly larger than in panel (a)]. This is further detailed 
in what follows. 

\begin{figure}[h]
\includegraphics[width=1.0\columnwidth]{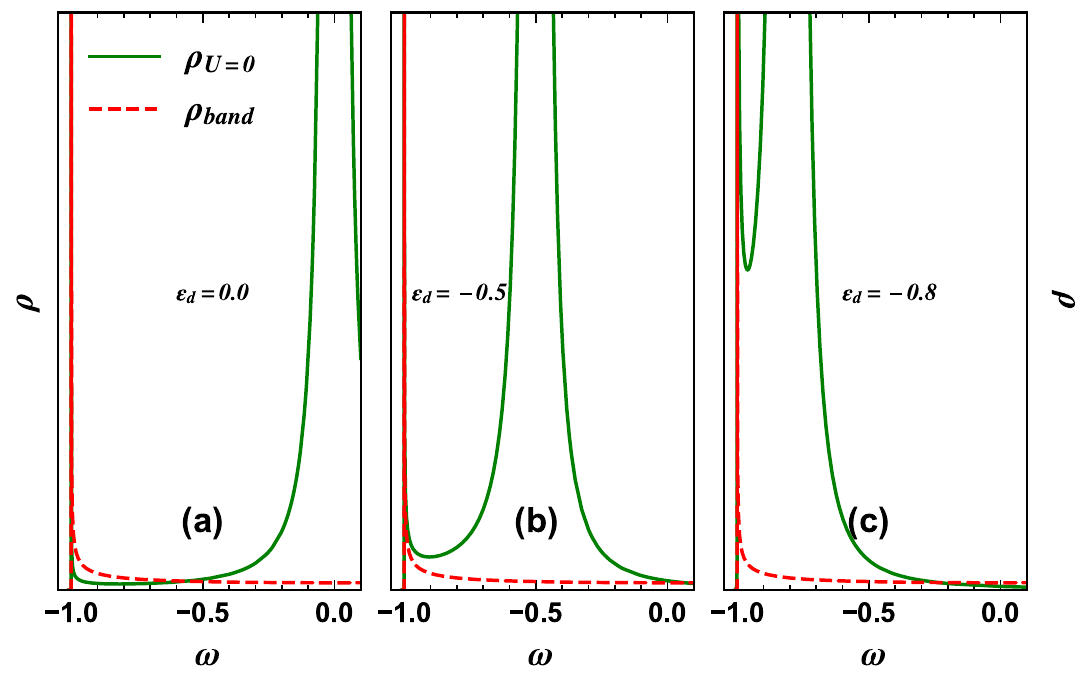}
	\caption{\label{fig13} Resonant level DOS $\rho_{U=0}$ (green curve)  
	for three different positions of the RL, $\e_d = 0.0$, 
	$-0.5$, and $-0.8$, in panels (a), (b), and (c), respectively.  
	Notice the large increase of the spectral weight of 
	the peak at the band singularity (red curve) as $\e_d$ approaches 
	the bottom of the band. $V=0.15$ in all panels. 
	} 
\end{figure}

In Fig.~\ref{fig12}, we show results for $Z_{b}$, which is defined as 
\begin{eqnarray}\label{zimp}
	Z_{b} = \int_{-\infty}^{\omega_{sing}} \rho_{U=0}(\omega)d\omega,
\end{eqnarray}
for the interval $0.0 \le \vert\gamma\vert \le 0.5$, where $\omega_{sing}$ is 
the position of the singularity. It clearly shows that 
$Z_b$ decreases monotonically with $\vert\gamma\vert$. This can be understood by 
analyzing the $\rho_{band}$ results in Fig.~\ref{fig10}, where we can easily 
see that for $\vert\gamma\vert=6.0$ there is considerably less spectral weight at the 
bottom of the band than for $\vert\gamma\vert=0$. Indeed, integrating $\rho_{band}$ 
from $\omega_{sing}$ to $\omega_{sing}+0.05$ we obtain $\approx 0.09$ 
for $\vert\gamma\vert=0$ and $0.04$ for $\vert\gamma\vert=6.0$. Thus, one expects that 
the RL will be less affected by the singularity for finite SOI. Again, this occurs 
because a finite SOI increases the bandwidth.

We have established that the presence of SOI in 1D 
weakens both the band-edge singularity and the 
resulting bound state. This is in contrast to the 
3D system in~\cite{Agarwala2016} that requires SOI 
to create a singular DOS. In what follows 
we analyze the properties of the 1D system without SOI.

\begin{figure}[h]
\center
\includegraphics[width=1.0\columnwidth]{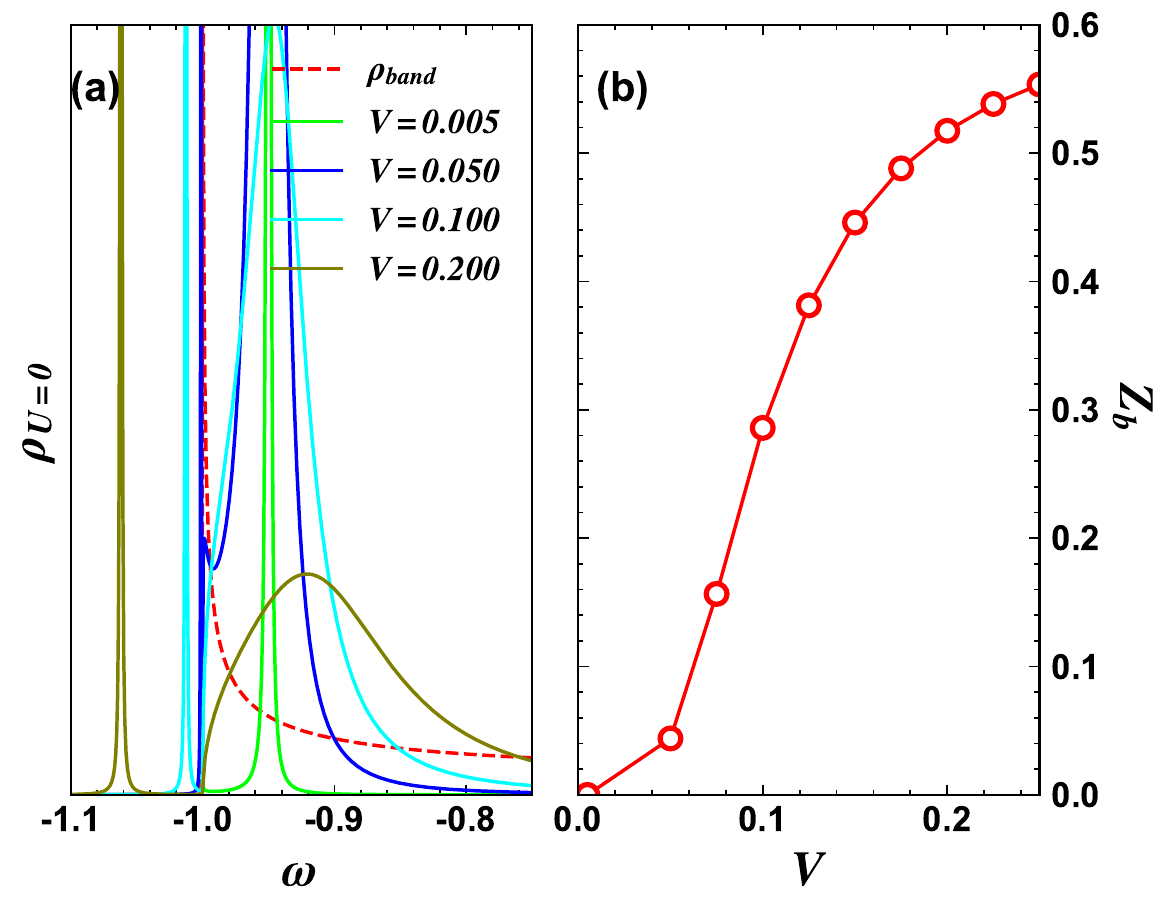}
	\caption{\label{fig14} (a) Variation of the RL DOS $\rho_{U=0}$ 
	with the coupling to the band in the interval $0.005 \le V \le 0.2$, 
	for $\vert\gamma\vert=0.0$ and $\delta=0.05$. (b) The variation of 
	$Z_b$ for $0.01 \le V \le 0.25$. 
	} 
\end{figure}

\section{Bound state properties}\label{appC}

\subsection{Bound state and coupling to the band}

First, we find that no matter what 
the energy of the RL is in relation to the singularity, there 
is always a bound state located either at the singularity or below it~\cite{dirac}.  
The latter occurs when the coupling of the RL to 
the band is strong, or if the RL is close to 
the singularity. In Fig.~\ref{fig13}, we show the RL DOS, $\rho_{U=0}$, for three different positions 
of the RL in relation to the bottom of the band. 
In panel (a), the RL is located at the center of the band, $\e_d=0$, and the singularity 
is at $\omega=-1.0$. We notice a vanishingly narrow DOS peak at the singularity. 
In panel (b), the RL is located at $\e_d=-0.5$, midway between the center of the band and the 
singularity. The DOS peak at the singularity has increased 
considerably. Finally, when the RL is just $0.2$ above the singularity,  
$\e_d=-0.8$, the bound state spectral weight at the 
singularity has increased drastically. If the 
coupling increases, and/or the RL approaches the singularity even more, 
the bound state detaches from the band continuum and moves to lower energies, 
as will be seen below.

\begin{figure}[h]
\center
\includegraphics[width=1.0\columnwidth]{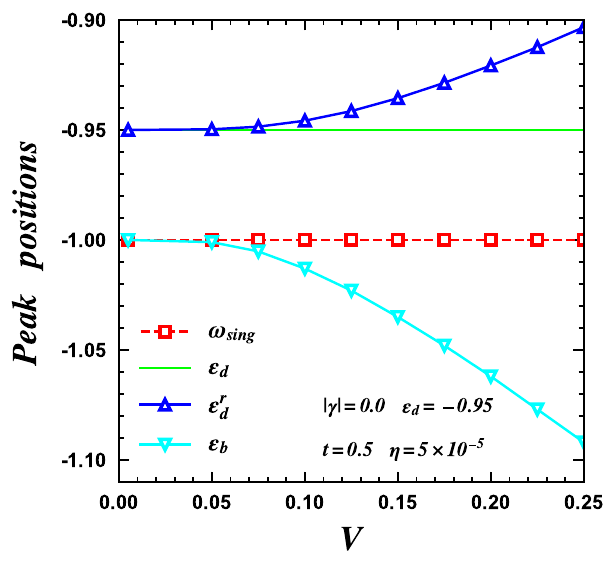}
	\caption{\label{fig15} Analysis of the formation of the bound state in the 
	RL spectral function $\rho_{U=0}$, 
	as a function of $V$ [obtained from the results in Fig.~\ref{fig14}(a)]. 
	Red curve shows the bottom of the band (position of the singularity), the green curve shows the position 
	of $\e_d$, the blue curve shows $\e_d^r$, the renormalized position of 
	the RL orbital energy, while the cyan curve shows the position 
	of the bound state, $\e_b$. 
	} 
\end{figure}

Fig.~\ref{fig14}(a) shows the RL DOS as 
we vary its coupling to the band in the range $0.005 \le V \le 0.2$, 
while keeping $\e_d = \omega_{sing} + 0.05$. For the smallest $V=0.005$ 
(green curve), a bound state at the singularity is not visible 
(vanishingly small spectral weight). For $V=0.05$ (blue curve) 
a very sharp peak at the singularity is already visible (with $Z_b=0.044$), 
while for $V=0.1$ (cyan curve) the bound state has detached from 
the bottom of the band and moved to lower energies. Its spectral weight 
has also increased to $Z_b=0.1566$. This trend continues as $V$ increases. 
Figure \ref{fig14}(b) shows the $Z_b$ increase with $V$, reaching 
more than half of the total spectral weight for $V=0.25$. 

Now, we analyze the data in Fig.~\ref{fig14} in more detail. 
Figure \ref{fig15} shows how the splitting of the RL $\e_d$ (green circles) 
into two parts, one, a renormalized peak $\e_d^r$ (blue up-triangles) inside the continuum, 
and the bound state $\e_b$ (magenta down-triangles), below the bottom of the band, 
progresses as one increases $V$. The red squares mark the bottom of the band. 

\begin{figure}[h]
\center
\includegraphics[width=1.0\columnwidth]{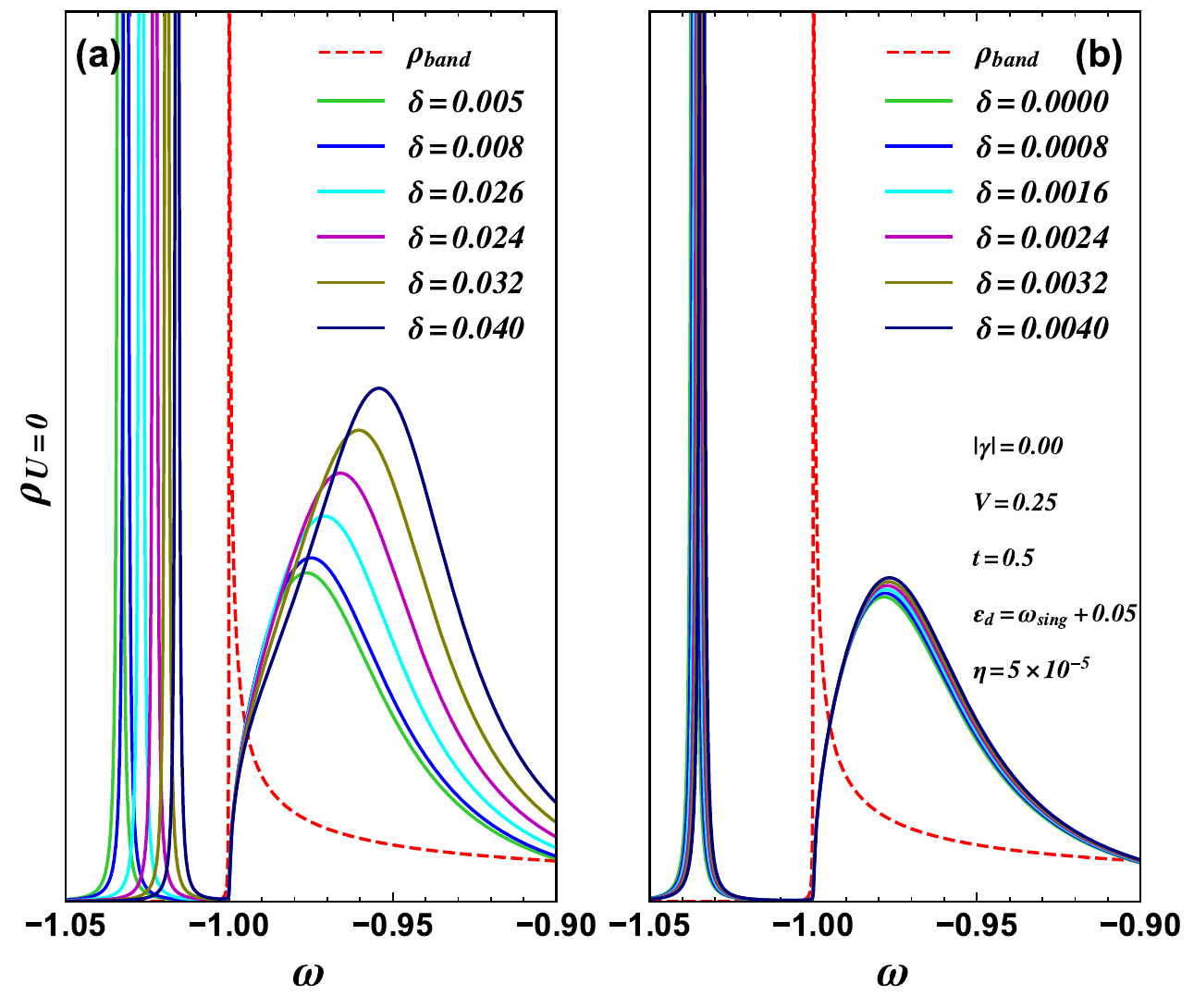}
	\caption{\label{fig16} Panels (a) and (b) show the variation of $\rho_{U=0}$ 
	with $\delta$ ($\e_d$ offset from the bottom of the band) for two different 
	intervals: (a) $0.005 \le \delta \le 0.04$ and (b) $0.0005 \le \delta \le 0.004$. 
	Panel (a) results show that, starting from $\delta=0.08$ (dark blue curve), 
	up to $\delta=0.016$ (blue curve), changes in $Z_b$ and peak positions 
	are considerable, while panel (b) results show that further approaching 
	$\e_d$ from the bottom of the band ($\delta \lesssim 0.01$) has limited 
	effects. Results obtained for $\vert\gamma\vert=0.0$ and $V=0.25$. 
	} 
\end{figure}

\begin{figure}[h]
\center
\includegraphics[width=1.0\columnwidth]{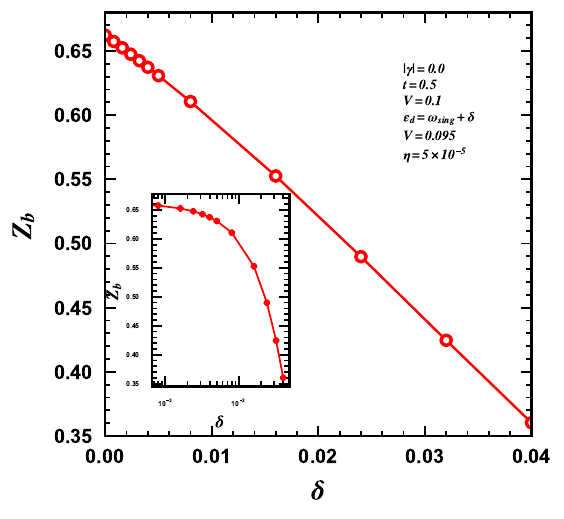}
	\caption{\label{fig17} Variation of the bound state spectral weight $Z_b$ 
	as a function of $\delta$ in the interval $0.001 \le \delta \le 0.08$. 
	Results obtained from both panels in Fig.~\ref{fig16}. The inset shows 
	the same results, but with a $\log$ scale in the $\delta$-axis, highlighting 
	the approach to the $Z_b=\nicefrac{2}{3}$ value. 
	} 
\end{figure}

\subsection{Bound state and distance to singularity}

Now, we analyze how the bound state varies as we move the RL $\e_d$ 
closer to the singularity. We set $\e_d=\omega_{sing} + \delta$, where $\omega_{sing}$ 
marks the bottom of the band, and vary $0.0 \le \delta \le 0.04$ (we fix $V=0.1$). 
The results are shown in Figs.~\ref{fig16}(a) and (b). Once 
$\e_d$ approaches $\omega_{sing}$ ($\delta$ tends to zero), the variation is very small, 
i.e., $\e_b$ and $Z_b$ tend to a fixed value. 

\begin{figure}[h]
\center
\includegraphics[width=1.0\columnwidth]{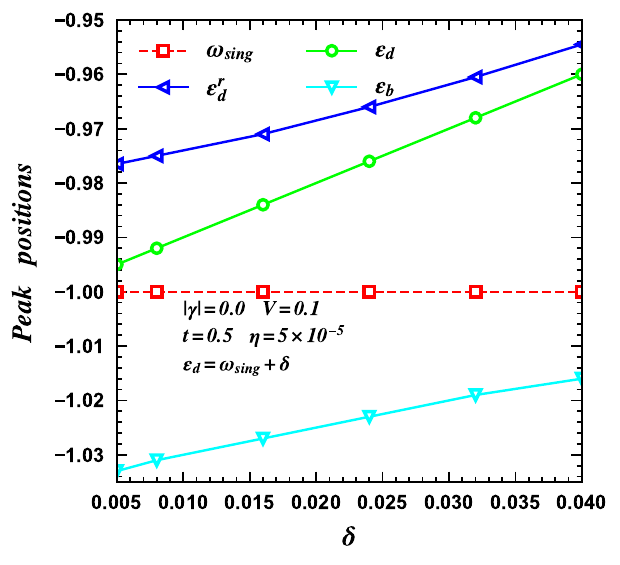}
	\caption{\label{fig18} Similar results as in Fig.~\ref{fig15}, 
	but now for the variation of $\delta$. The results shown are just 
	for panel (a) in Fig.~\ref{fig16}. 
	} 
\end{figure}

\begin{figure}[h]
\center
\includegraphics[width=1.0\columnwidth]{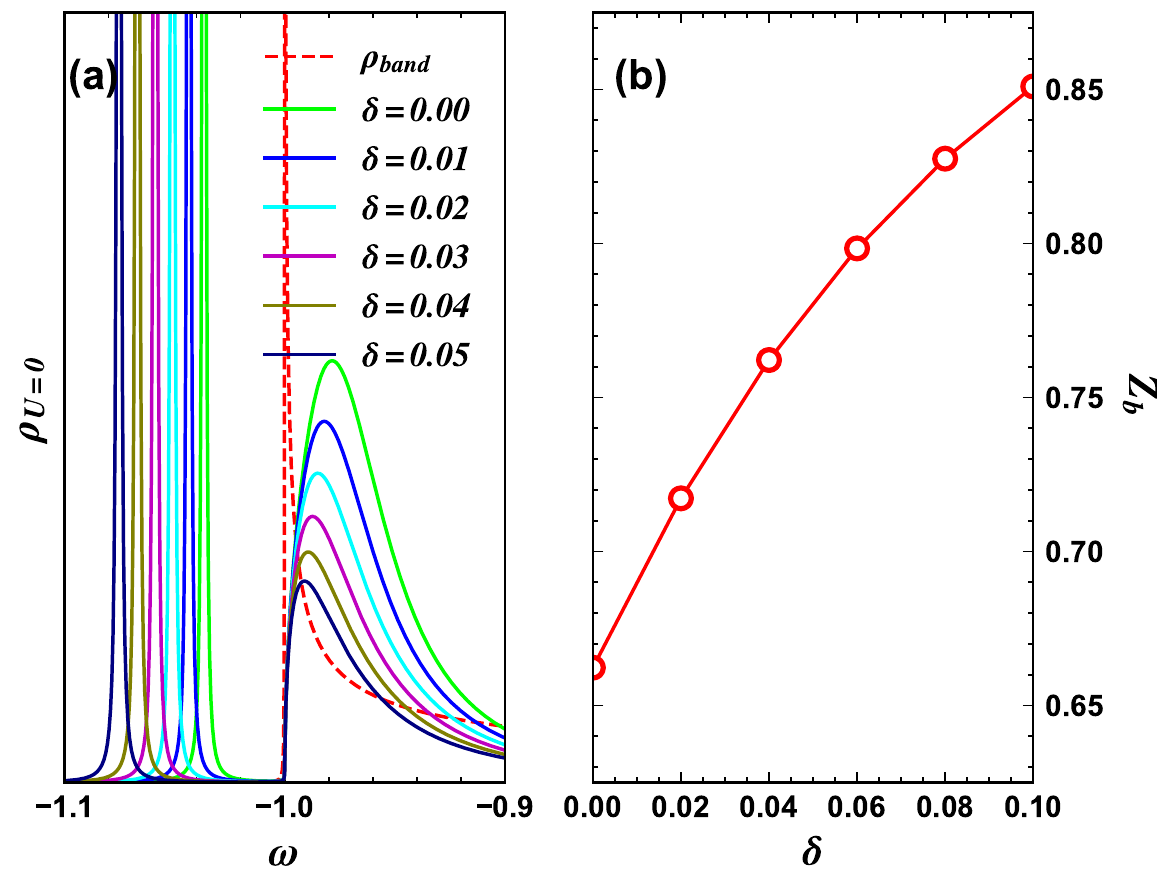}
	\caption{\label{fig19} Similar results as in Fig~\ref{fig16}, 
	but now when the impurity is placed below the bottom of the band 
	(note that $\e_d=\omega_{sing} - \delta$) and $V=0.1$. 
	(a) Results for $\rho_{U=0}$, for $0.0 \le \delta \le 0.1$, 
	showing again the peak splitting. (b) Spectral weight $Z_b$ of the bound state, 
	now larger than $\nicefrac{2}{3}$. 
	} 
\end{figure}

The variation of $Z_b$ with $\delta$ may be seen in Fig.~\ref{fig17}. 
We see that $\delta \rightarrow 0.0$ implies 
$Z_b \rightarrow \nicefrac{2}{3}$, in agreement with the results 
obtained in Ref.~\cite{Agarwala2016}. This shows the very interesting 
phenomenon that the $Z_b=\nicefrac{2}{3}$ result does not depend on the details 
of the band, such as spatial dimensionality (3D vs 1D) and presence vs 
absence of SOI. The inset, with a $\log$ scale in the $\delta$-axis, 
emphasizes the gradual approach to the $Z_b=\nicefrac{2}{3}$ value.

\begin{figure}[h]
\center
\includegraphics[width=1.0\columnwidth]{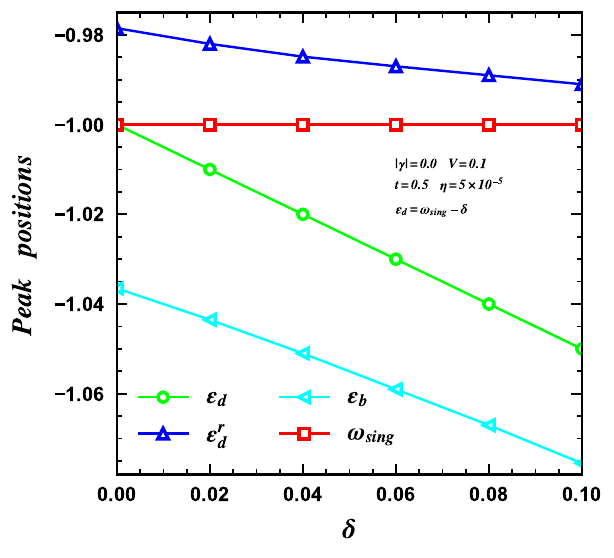}
	\caption{\label{fig20} Similar results as in Fig.~\ref{fig18}, 
	but now $\e_d$ (green curve) is placed below the 
	bottom of the band. Larger peak separations are obtained this 
	way. 
	} 
\end{figure}

Figure \ref{fig18} shows details of the results in Fig.~\ref{fig16}(a), 
as done in Fig.~\ref{fig15} for the results in Fig.~\ref{fig14}(a). 

Figure \ref{fig19} shows what happens when we place $\e_d$ out of the 
continuum, i.e., $\e_d = \omega_{sing} - \delta$ and $0.0 \le \delta \le 0.05$, 
with $V=0.1$. Panel (b) shows that $Z_b$ takes values above $\nicefrac{2}{3}$, 
increasing with $\delta$. The DOS inside the continuum, as can be seen 
in panel (b), tends to accumulate at the bottom of the band as $\delta$ increases. 

Finally, Fig.~\ref{fig20} shows details of the results in Fig.~\ref{fig19}(a).

\section{Study of the impurity spectral function}\label{appD}

\begin{figure}[h]
\center
\includegraphics[width=1.0\columnwidth]{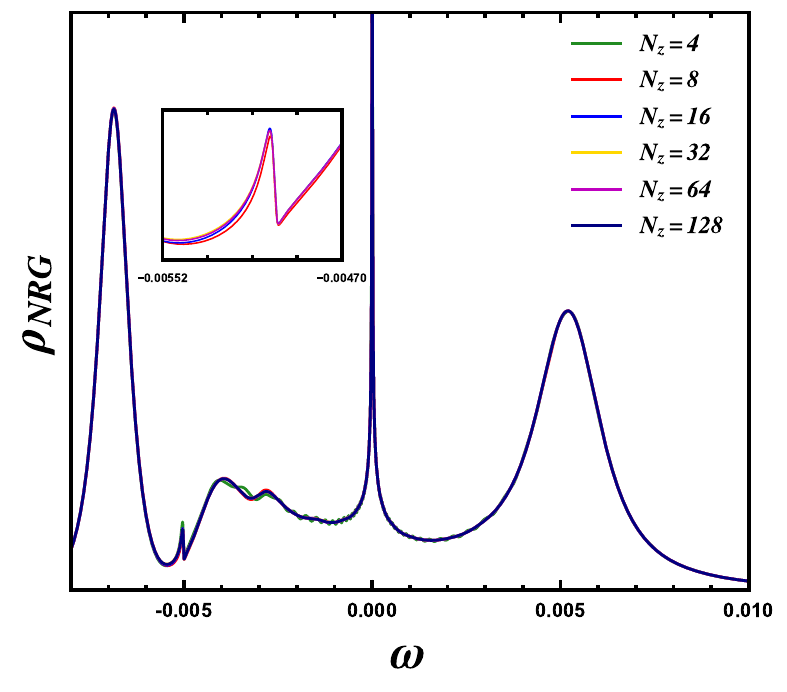}
	\caption{\label{fig21} Analysis of the $\rho_{NRG}$ dependence on 
	$N_z$ (interleaving parameter~\cite{Zitko2009c}) for the results 
	in Fig.~\ref{fig2}(b), in the interval $4 \leq N_z \leq 128$. 
	The inset shows a zoom in the region where the curves 
	(except for $N_z=4$) vary the most (around $P_1$). 
	Taking the peak-height change as a measure of the overall variation in this region, 
	we obtain that the variation between $N_z=16$ and $N_z=8$, $N_z=32$ and $N_z=16$, 
	$N_z=64$ and $N_z=32$, and $N_z=128$ and $N_z=64$, is $4.1$\%, $1.5$\%, 
	$0.4$\%, and $0.08$\%, respectively. 
	Thus, for the purpose of presentation, the results are already well-converged 
	for $N_z=16$ (value used in this work). As to exposing numerical artifacts, good 
	convergence has been achieved for $N_z=32$ (see text).
	} 
\end{figure}

In this section, we follow Ref.~\cite{Zitko2009c} and do an step by step analysis of the results 
in Figs.~\ref{fig2} and \ref{fig4}, to show that the NRG spectral function 
features discussed in Secs.~\ref{sec-spectral} and \ref{sec-correl} are not 
numerical artifacts. There are quite a few aspects that should be taken into account 
to root out NRG numerical artifacts in the impurity spectral function $\rho_{NRG}$. 
According to \v{Z}itko and Pruschke~\cite{Zitko2009c}, overbroadenig effects 
reduce energy resolution at higher energies and wash out spectral features with 
small spectral weight (like $P_1$, in Figs.~\ref{fig2} and \ref{fig4}, for the larger values of $U/\Gamma$). 
The so-called interleaved method (or `$z$ averaging')~\cite{Campo2005} allows for the use of 
narrower broadening functions, mitigating overbroadening and removing 
oscillatory features in the impurity spectral function. The method consists of performing 
several ($N_z$) NRG calculations for different logarithmic discretization 
meshes and then taking their average to obtain the final impurity spectral function. 
However, one has to check that convergence has been attained, before 
trying to root out artifacts. Figure \ref{fig21} shows the evolution 
of $\rho_{NRG}$ with increasing $N_z$ in the interval $4 \leq N_z \leq 128$. 
It is clear that $N_z=4$ (green curve) is not nearly enough; however, 
for $N_z \geq 16$, $\rho_{NRG}$ has converged. It is interesting to note 
that $P_1$ is the last feature to converge. Indeed, the largest difference (which 
occurs around $P_1$, as shown in the inset) between $N_z=16$ and $8$ is $\approx 4.1$\%, 
while it is $\approx 1.5$\% for the difference between $N_z=32$ and $16$. 
This value falls to $\approx 0.4$\% for $N_z=64$ (in relation to $32$), 
and to $\approx 0.08$\% for $N_z=128$ (in relation to $64$). Thus, 
for the purpose of exposing numerical artifacts~\cite{Zitko2009c} (see next step), 
all spectral function features are well-converged for $N_z=32$.

\begin{figure}[h]
\center
\includegraphics[width=0.8\columnwidth]{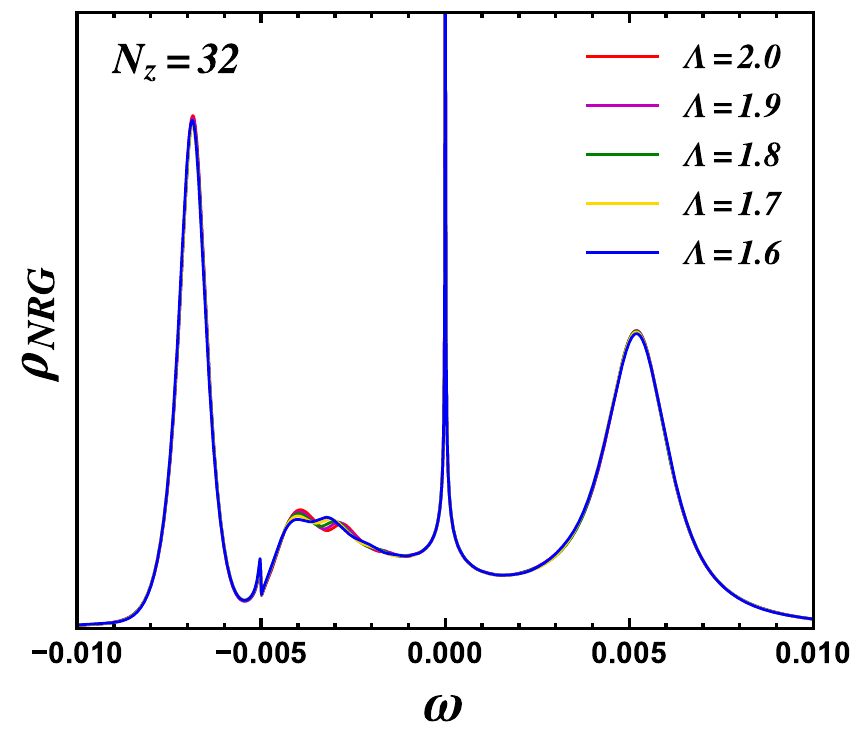}
	\caption{\label{fig22} Analysis of the $\Lambda$ dependence of the 
	$\rho_{NRG}$ results in Fig.~\ref{fig2}(b) in the interval 
	$1.6 \leq \Lambda \leq 2$. As shown in Ref.~\cite{Zitko2009c}, 
	NRG discretization artifacts should shift position and 
	change form significantly with decreasing $\Lambda$. 
	Note that we use the well-converged $N_z=32$ results for this analysis. 
	} 
\end{figure}

The next step, once $N_z$ convergence of all $\rho_{NRG}$ features has been ensured, is to analyze 
its dependence with $\Lambda$ (discretization parameter). 
Indeed, artifacts will shift (and change form) substantially when $\Lambda$ 
varies, while real features will change very little~\cite{Zitko2009c}. 
Usually, the detection of artifacts is more effectively done when one studies 
the approach to the continuum limit ($\Lambda \rightarrow 1$). Thus, 
we decreased the discretization parameter below the standard $\Lambda=2$ 
value (down to $1.6$). Figure \ref{fig22} shows $\rho_{NRG}$ results 
for the same parameters as in Fig.~\ref{fig2}(b) (except that now $N_z=32$), for 
the interval $1.6 \leq \Lambda \leq 2$ (varying in steps of $0.1$). It is clear 
that peaks $P_0$ and $P_1$, which were discussed in detail in Sec.~\ref{sec-correl}, 
suffer marginal changes, indicating that they are not numerical artifacts.

\begin{figure}[h]
\center
\includegraphics[width=1.0\columnwidth]{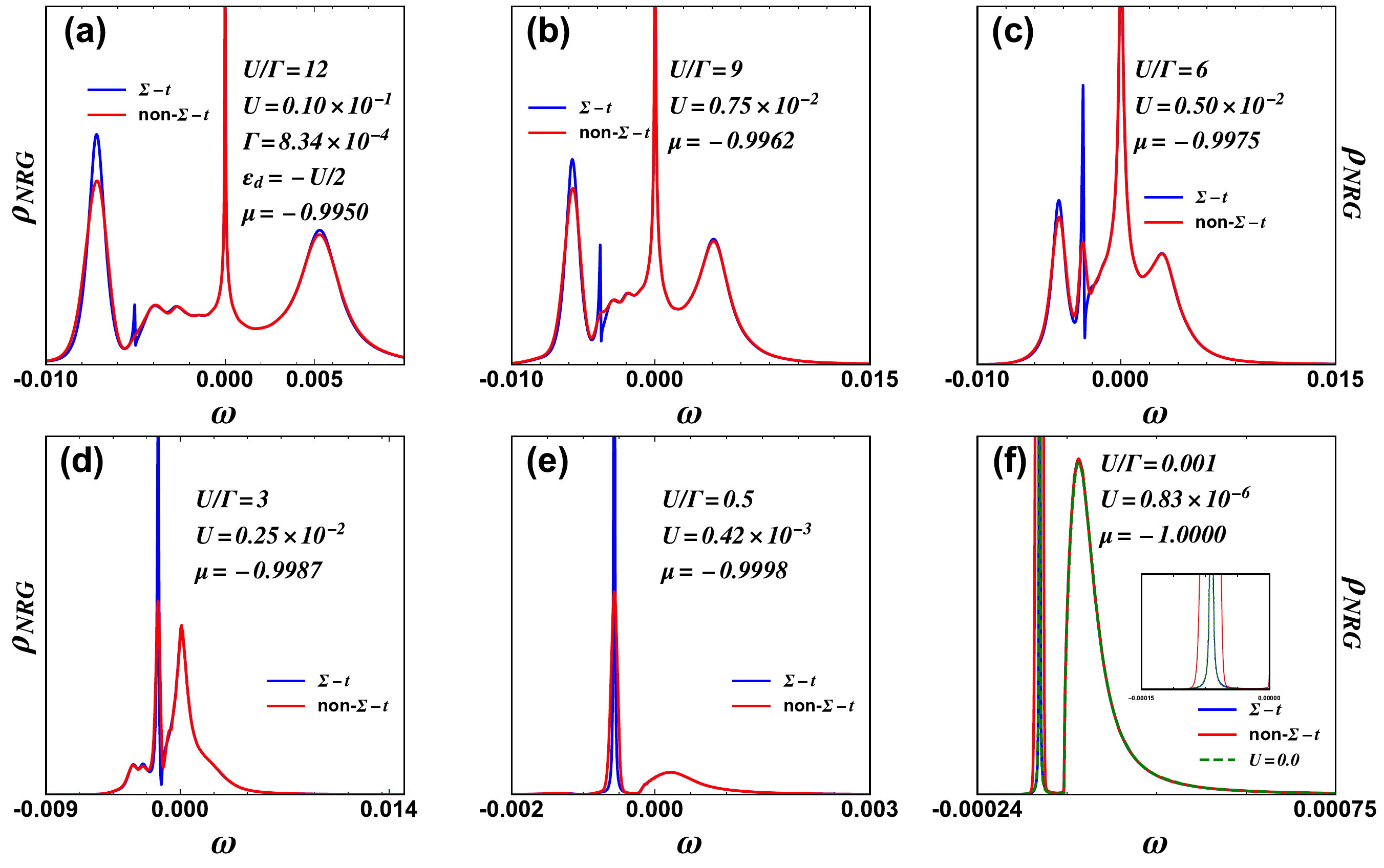}
	\caption{\label{fig23} Reproduction of the results in Fig.~\ref{fig4}, 
	comparing the impurity spectral function $\rho_{NRG}$ calculated using the 
	so-called `self-energy trick'~\cite{Bulla1998} (blue curves) with the results 
	obtained without its use (red curves). The inset in panel (f) zooms in 
	on the bound state, showing that the self-energy trick removes the overbroadening 
	and produces the exact result (dashed green curve). 
	} 
\end{figure}

One last step consists in applying the so-called `self-energy trick'~\cite{Bulla1998} 
(dubbed $\Sigma$-t, for short, in Fig.~\ref{fig23}), 
which is a very efficient method to reduce overbroadening effects~\cite{Zitko2009c}. This `trick' consists 
in calculating the impurity self-energy as the ratio of two correlation functions, and then 
using it to obtain the impurity Green's function, whose imaginary part is proportional to the impurity 
spectral function. It is important to remark that all our spectral function results in this work were 
obtained by using $\Sigma$-t. In Fig.~\ref{fig23}, we compare our results 
for Fig.~\ref{fig4} (obtained through $\Sigma$-t, blue curves) with the results obtained without 
the use of $\Sigma$-t (red curves). There are some interesting points to stress. First, in all panels, 
the use of $\Sigma$-t results in the narrowing of some features, most notably of $P_1$, mainly in panels (a) 
and (b), where $P_1$ is barely noticeable without $\Sigma$-t [especially in panel (a)]. 
In addition, we want to call special attention to the result in panel (f), where the use of 
$\Sigma$-t (blue curve) has narrowed the non-$\Sigma$-t result (red curve) almost perfectly  
into the exact non-interacting ($U=0$) result (dashed green curve). This is emphasized in 
panel (f)'s inset. Since there is no reason for the accuracy of the $\Sigma$-t spectral function 
to be reduced for finite $U$~\cite{Bulla1998}, we can have confidence in the accuracy 
of the spectral function results presented in Figs.~\ref{fig2} and \ref{fig4}, 
in the main text.

The NRG package used here (NRG Ljubljana~\cite{zitko_rok}) has the so-called 
`patching procedure'~\cite{Bulla2001} fully implemented, and it was used in all 
spectral function calculations done here. It is well-known that the NRG is an 
iterative procedure which solves consecutive so-called Wilson chains with increasing sizes $N$. 
The energy window being analyzed, at each specific stage of the iterative 
procedure, logarithmically approaches the ground state region of the spectra  
for two consecutive iterations. One has to carefully `join' (patch) 
the spectral information acquired at iterations $N+2$ and $N$. This 
procedure results in a smooth spectral function across energy windows 
at different energy scales.

\begin{figure}[h]
\center
\vspace*{0.5cm}
\includegraphics[width=1.0\columnwidth]{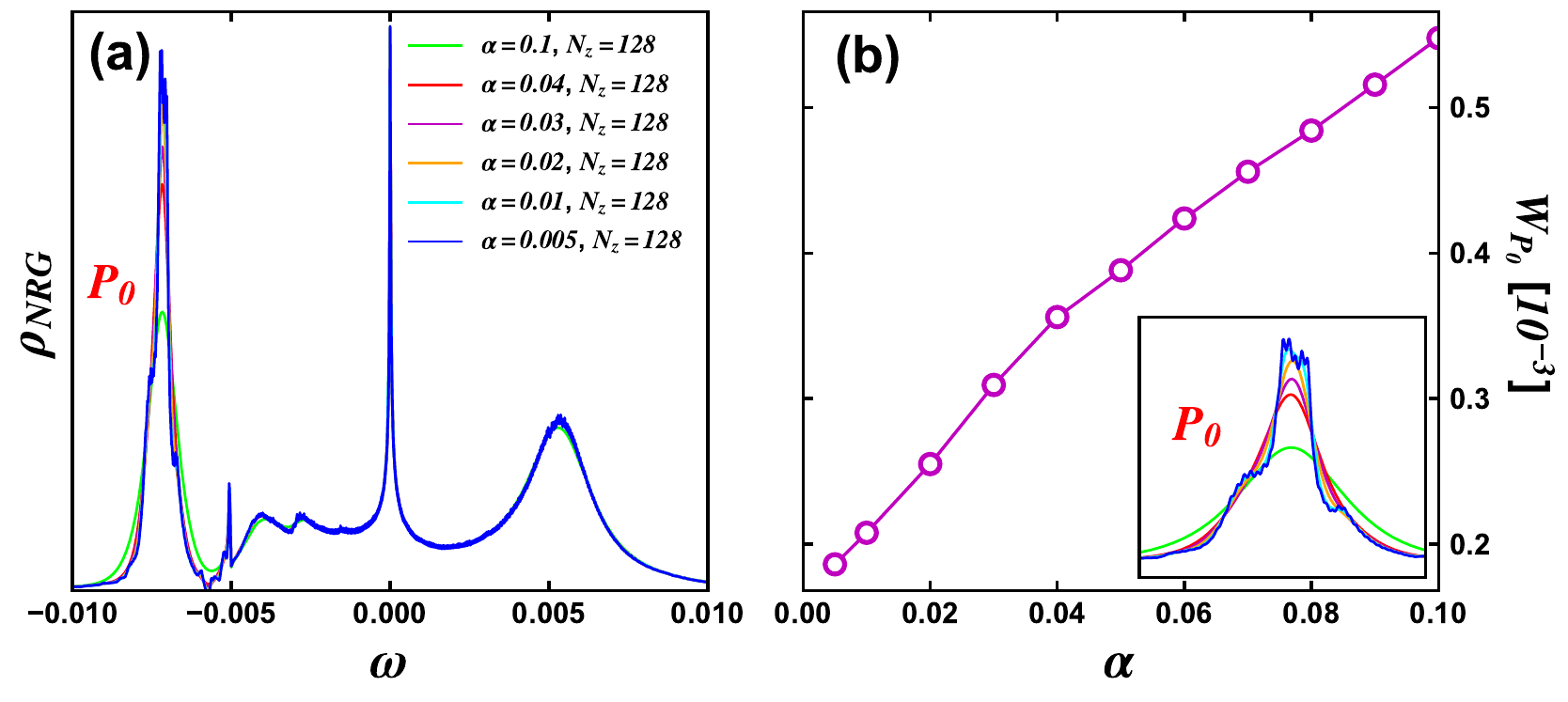}
	\caption{\label{fig24} (a) Variation of $\rho_{NRG}$---same parameters 
	as in Fig.~\ref{fig4}(a), but now for $N_z=128$ 
	instead of $16$---with the NRG broadening parameter $0.005 \leq \alpha \leq 0.1$. 
	(b) Half-width at half-height $W_{P_0}$ of the $P_0$ peak in panel (a) 
	as a function of $\alpha$. The inset shows a zoom of $P_0$. 
	} 
\end{figure}

Finally, to analyze if the finite-width of $P_0$, observed in Fig.~\ref{fig4}, is 
caused by NRG overbroadening, we present, in Fig.~\ref{fig24}(a), $\rho_{NRG}$  
for the same parameters as in Fig.~\ref{fig4}(a), but now for $N_z=128$ (instead of 16), 
and varying values of the broadening NRG parameter 
$0.005 \leq \alpha \leq 0.1$~\footnote{The Gaussian broadening 
used, $\eta$, is related to $\alpha$, by $\eta = \alpha |E|$, where $E$ is the 
pole being broadened.}. As shown in Fig.~\ref{fig21}, the result for $\alpha=0.1$ 
(green curve) [same $\alpha$ as used in Fig.~\ref{fig4}(a)], changes very little 
when we increase $N_z$, in this case, from $N_z=16$ [Fig.~\ref{fig4}(a), blue curve] 
to $N_z=128$ [Fig.~\ref{fig24}(a), green curve]. For decreasing $\alpha$ values, 
we see that the largest changes in $\rho_{NRG}$ occur for $P_0$ [a zoom of $P_0$ 
is shown in the inset in Fig.~\ref{fig24}(b)]. As shown in panel (b), 
the $P_0$ half-width at half-height, denoted $W_{P_0}$, decreases 
by $67$\% (from $5.5 \times 10^{-4}$ to 
$1.8 \times 10^{-4}$), while $\alpha$ decreases $20$ times (from $0.1$ to $0.05$). 
However, one cannot rule out the possibility that the peak width will 
extrapolate to zero. Thus, NRG cannot conclusively resolve this issue.

\bibliography{sing-Kondo}

\begin{thebibliography}{60}
\bibitem [{che()}]{chempot}%
  \BibitemOpen
  \href@noop {} {}\bibinfo {note} {The sign of the chemical potential in
  Eq.~\eqref{h1D} is such that $\mu=0$ corresponds to half-filling (Fermi
  energy at the center of the band), and $\mu=-D=-1.0$ corresponds to the Fermi
  energy at the singularity (at the bottom of the band).}\BibitemShut {Stop}%
\bibitem [{edg()}]{edge}%
  \BibitemOpen
  \href@noop {} {}\bibinfo {note} {Note that the spectral weight of the
  bound-state (not shown) for $\mu=0$ is vanishingly small (see
  Appendix).}\BibitemShut {Stop}%
\bibitem [{fin()}]{finitewidth}%
  \BibitemOpen
  \href@noop {} {}\bibinfo {note} {In Appendix~\ref{appD}, we were able to show
  that most of the features present in the impurity spectral function are real.
  Although suggestive of an actual finite width for the bound state (peak
  $P_0$), one cannot ascertain its finite value and independence of calculation
  parameters.}\BibitemShut {Stop}%
\bibitem [{kon()}]{kondotemp}%
  \BibitemOpen
  \href@noop {} {}\bibinfo {note} {Notice that at the bottom of
  Fig.~\ref{fig3}(a) the system is in an intermediate valence regime, so that a
  proper Kondo temperature is not well defined. However, as will be shown in
  Figs.~\ref{fig6} and \ref{fig7}, the presence of the bound state strongly
  affects the LM fixed point of the impurity, suppressing its charge
  fluctuations, forcing it back into the Kondo regime.}\BibitemShut {Stop}%
\bibitem [{dir()}]{dirac}%
  \BibitemOpen
  \href@noop {} {}\bibinfo {note} {We have checked that the $U=0$ bound state
  is a Dirac delta function, since its half-width at half-height is exactly
  $\eta$ [see Eq.~\eqref{Gimp}].}\BibitemShut {Stop}%
\end{thebibliography}

\end{document}